\documentclass[12pt,a4paper]{article}
\usepackage{graphicx}
\usepackage{cite}

\makeatletter
\renewcommand{\@biblabel}[1]{}
\makeatother

\begin{document}
	
\setlength{\baselineskip}{0.77cm}

\newcommand{\EQ}{equation~}
\newcommand{\EQS}{equations~}
\newcommand{\FIG}{figure~}
\newcommand{\FIGS}{figures~}
\newcommand{\SEC}{section~}
\newcommand{\SECS}{sections~}

\begin{center}
{\Huge Formation of Feedforward Networks and Frequency Synchrony by Spike-timing-dependent Plasticity}\\
\bigskip
{\bf Naoki Masuda}\\
Amari Research Unit,
RIKEN Brain Science Institute,\\
2-1 Hirosawa, Wako, Saitama 351-0198 Japan\\
\bigskip
{\bf Hiroshi Kori}\\
Department of Mathematics, Hokkaido University,\\
Kita 10, Nishi 8, Kita-Ku, Sapporo, Hokkaido, 060-0810 Japan
\end{center}

\bigskip
\bigskip
\bigskip
\bigskip
\bigskip

\begin{flushleft}
Corresponding author: Naoki Masuda\\
Tel: +81-48-467-9664\\
Fax: +81-48-467-9693\\
Email: masuda@mist.i.u-tokyo.ac.jp
\end{flushleft}

\newpage

\begin{center}
{\bf Abstract}
\end{center}

Spike-timing-dependent plasticity (STDP) with asymmetric learning windows is commonly found in the brain and useful for a variety of spike-based computations such as input filtering and associative memory. A natural consequence of STDP is establishment of causality in the sense that a neuron learns to fire with a lag after specific presynaptic neurons have fired. The effect of STDP on synchrony is elusive because spike synchrony implies unitary spike events of different neurons rather than a causal delayed relationship between neurons. We explore how synchrony can be facilitated by STDP in oscillator networks with a pacemaker. We show that STDP with asymmetric learning windows leads to self-organization of feedforward networks starting from the pacemaker. As a result, STDP drastically facilitates frequency synchrony. Even though differences in spike times are lessened as a result of synaptic plasticity, the finite time lag remains so that perfect spike synchrony is not realized. In contrast to traditional mechanisms of large-scale synchrony based on mutual interaction of coupled neurons, the route to synchrony discovered here is enslavement of downstream neurons by upstream ones. Facilitation of such feedforward synchrony does not occur for STDP with symmetric
learning windows.

\bigskip
\bigskip

Keywords: spike-timing-dependent plasticity, synchronization, feedforward networks, complex networks

\newpage

\section{Introduction}\label{sec:introduction}

In many neural circuits, synaptic plasticity depends on relative
timing of presynaptic and postsynaptic spikes, which is known as
spike-time dependent plasticity (STDP)
\cite{Gerstner96nat,Bell97,Markram97,Bi98,Zhang98}.
Specifically, long-term potentiation (LTP)
ensues when a presynaptic neuron fires slightly
before a postsynaptic neuron (of
the order of 10 ms), whereas long-term
depression (LTD) is
elicited in the opposite case (solid line in \FIG\ref{fig:windows}).
This STDP rule promotes causal
relationship between a pair of neurons in the sense that the strength
of a synapse that contributes to generation of
postsynaptic spikes is reinforced.  Computationally, STDP is
useful for synaptic competition
\cite{Kempter99,Song00,Vanrossum00,Song01}, coincidence detection
\cite{Gerstner96nat}, spike-based associative
memory (e.g. Lengyel et al., 2005),
implementation of the synfire chain
\cite{Horn,Levy}, generation of reproducible spatiotemporal spike
patterns \cite{Izhikevich04CC,Izhikevich06NC}, selection of earlier
inputs, to name a few 
(e.g. Gerstner and Kistler, 2002).

Related to neural computation, coincident firing of multiple neurons
in the oscillatory regime is found in many parts of the brain and
believed to play an important role \cite{Singer95,Ritz97,Buzsaki04}.
One could
imagine that real neural networks learn to synchronize spikes of
different neurons by STDP-related synaptic plasticity, as suggested
by some modeling studies. However,
contribution of STDP to spike synchrony may be limited.  For example,
STDP can lead to division of a neural population into
clusters in each of which neurons fire in spike synchrony \cite{Horn,Levy}.
This self-organizing process actually
necessitates homogeneous synaptic transmission delays 
for different synapses and puts a strong restriction on the firing
period. If there is just one cluster (all
neurons firing synchronously), the firing period has to be equal to
the synaptic transmission delay.
Similarly, if there are two clusters, the first cluster
excites a synchronous volley in the second cluster after the synaptic
transmission delay. The second cluster reexcites the first cluster
in a similar way. The firing period is equal to
the synaptic transmission delay multiplied by the number of
clusters, which seems
restrictive. Alternatively, coincident firing is achieved
via STDP if the amount of LTP and that of LTD that are
caused by a presynaptic and postsynaptic spike pair
are perfectly balanced \cite{Karbowski}. Evolution of coincident
firing survives heterogeneity in neurons and in the amount of
plasticity. However, how coincident firing
 is affected by the imbalance between LTP
and LTD remains to be explored.

Coincident firing
in recurrent neural networks may not be established through STDP. In
general, synchronous firing can be induced by sufficiently strong coupling
between elements \cite{Kuramotobook,Pikovskybook,Gerstnerbook}. By
contrast, STDP cannot strengthen the synaptic weights between two
neurons bidirectionally. An increase of the synaptic weight in one
direction implies a decrease in the opposite direction, and stability
requires that the net decrease and the net increase are roughly balanced
\cite{Song00,Song01}. Therefore, STDP does not necessarily enhance
mutual interaction.  Indeed, in
recurrent neural networks, STDP does not necessarily support
synchronous firing \cite{MasudaSTDP}. It rather reinforces reproducible
spatiotemporal spike patterns composed of causal spike pairs of
different neurons \cite{Izhikevich04CC,Izhikevich06NC}.

We examine possible mechanisms of STDP-induced synchrony in recurrent
networks of oscillatory elements. We distinguish two
types of synchrony using the terminology of coupled
oscillators. One type is phase synchrony, which is equivalent to spike
synchrony. When neurons are in phase synchrony, they share 
spike timing.  The
other weaker notion is frequency synchrony in which
neurons possibly
with different intrinsic
firing rates share a common firing rate. Frequency
synchrony does not imply phase synchrony.
The spike time of the
postsynaptic neuron can differ from that of the presynaptic neuron.
According to these definitions, the previous studies
cited above, which relate STDP to synchrony, regard phase synchrony.

STDP may be more relevant to frequency synchrony.  For example, STDP
promotes frequency synchrony, but not phase synchrony, in a hybrid
circuit of an aplysia abdominal neuron and an emulated neuron
\cite{Nowotny03JNS}.  Unidirectional connectivity from the emulated
neuron to the aplysia neuron eventually forms. Numerical
simulations of two coupled Hodgkin-Huxley neurons \cite{Zhigulin03}
and of large neural networks \cite{Zhigulin04} also support the notion that
frequency synchrony is facilitated by STDP.

In the present work, we show that the standard STDP facilitates
frequency synchrony to a great extent, particularly when LTP
and LTD are roughly balanced.  To examine how heterogeneous
neurons interact to produce
possible synchronization, we analyze networks of
oscillators with a pacemaker. The pacemaker has a distinct natural
frequency, is not affected by other oscillators, and sets the rhythm
to influence other neurons.  No matter whether a pacemaker is realized by
a network or a single neuron, existence of pacemaker neurons is
suggested in, for example, the basal ganglia \cite{Plenz99} and
respiratory networks in the pre-B\"{o}tzinger complex
\cite{Ramirez04}.
Furthermore, many neurons \cite{Hutcheon00} and
recurrent microcircuits \cite{Jefferys} are intrinsically oscillatory
(also see e.g. Singer and Gray, 1995).
and their rhythmic activities are resistant to perturbation.  These
neural networks and single neurons can also serve as pacemakers.  With
frozen and sufficiently strong synapses, oscillator
networks with pacemakers allow frequency synchrony \cite{Kori04}. We
show that STDP considerably facilitates frequency synchrony of pacemaker
systems by establishing feedforward network structure 
whose root is the pacemaker.

For analytical tractability, we mostly deal with networks of phase
oscillators in which coupling strength evolves according to
STDP. Coupled phase oscillators approximate various natural systems
composed of self-sustained oscillators
with weak coupling \cite{Winfree80,Kuramotobook,Glassbook},
including pulse-coupled neurons
\cite{Kuramoto91,Hansel93,Hansel95,Kori03}. We introduce the model in
\SEC\ref{sec:model} and analyze simple cases of two connected
oscillators in \SEC\ref{sec:small}. We numerically analyze larger
oscillator networks with STDP in \SEC\ref{sec:large}.  In
\SEC\ref{sec:iz}, we numerically simulate pulse-coupled pyramidal neuron
models to show that our results obtained for coupled phase oscillators
qualitatively apply to
spiking neuron models.

\section{Model}\label{sec:model}

We analyze a network of $n$ phase oscillators.
One oscillator is assumed not to be disturbed by the other
$n-1$ oscillators. We designate this special oscillator as pacemaker
and use the term oscillator to refer to the other $n-1$ elements. The
pacemaker has natural frequency $\Omega$ and phase $\phi_0\in
[0,2\pi)$.  The other
oscillators are assumed to have the identical natural frequency
$\omega$, and the phase of the $i$-th oscillator
$(1\le i\le n-1)$ is denoted by $\phi_i \in [0,2\pi)$.
We identify $\phi_i=0$ and $\phi_i=2\pi$ ($0\le i\le n-1$).
We write
$(i,j)\in E$ if there is a synaptic connection from oscillator $i$ to
oscillator $j$. In other words, $E$ is the set of edges of the
underlying neural network.  As in real neural networks,
connectivity is asymmetric in general so that $(i,j)\in E$ does not
imply $(j,i)\in E$. A pair of connected oscillators
interact via sinusoidal coupling, which usually
promotes synchrony \cite{Kuramotobook}.

Dynamics for fixed synaptic strengths
are represented by:
\begin{eqnarray}
\dot{\phi}_0 &=& \Omega,\\
\dot{\phi}_i &=& \omega + \frac{1}{\left<k\right>}\sum_{j: (j,i)\in E} 
g_{ji}\sin(\phi_j - \phi_i),
\quad (1\le i\le n-1)
\label{eq:homo-oscillators}
\end{eqnarray}
where $\left<k\right>$ is the
average number of incoming edges per oscillator.
The coupling strength $g_{ji}$ is associated with synapse $(j,i)$.
We note that $g_{i0}$, which is the synaptic weight from an
oscillator to the pacemaker, does not affect the network dynamics: the
pacemaker is not perturbed by external input. However, we will monitor
$g_{i0}$ to examine how this connection evolves as synaptic
plasticity goes on.

We assume that $\Omega > \omega$ unless otherwise stated.  By rescaling the
timescale and the coupling strength, we set $\Omega = \omega +1$
without losing generality.  To set the values of $\Omega$ and
$\omega$, we take care of two subtle factors.  First, a small $\omega$
would yield backward rotation by the effect of coupling.
This is because the second term of the right-hand side of 
\EQ(\ref{eq:homo-oscillators}) can be large negative to overwhelm the
first term. Then, the condition $\dot{\phi}_i<0$ may be satisfied for
long enough time to elicit backward firing.
This is unrealistic as a neuron model.  Second, we avoid a pair of
$\Omega$ and $\omega$ that accommodates the relation $M_1 \Omega = M_2
\omega$ with small integers $M_1$ and $M_2 (M_1 \neq M_2)$. In such a 
situation, resonant behavior appears when the pacemaker and
the oscillators are decoupled through STDP and has a pathological effect
(see the explanation after \EQ(\ref{eq:a-dotgf}) for more details).
The resonant firing is ruled out by
dynamical noise in many real neural networks. 
However, we have to carefully
specify $\Omega$ and $\omega$ in the present work
because we do not assume noise for
analytical tractability.
Keeping these caveats in mind, we set $\Omega=9.1$
and $\omega =8.1$.

Spike time is defined to be the time when the $\phi_i$ crosses $0$.
Synaptic update based on STDP takes place based on a pair of nearest
presynaptic and postsynaptic spike times, without paying attention to
remote spike pairs
(see arguments in e.g. Froemke and Dan, 2002).
We compare the upshot of two types of STDP
rules for synapse $g_{ji}$ $((j,i)\in E)$,
namely,
{\it asymmetric} STDP and {\it symmetric} STDP.

\begin{figure}
\begin{center}
\includegraphics[height=6.0cm,width=6.0cm]{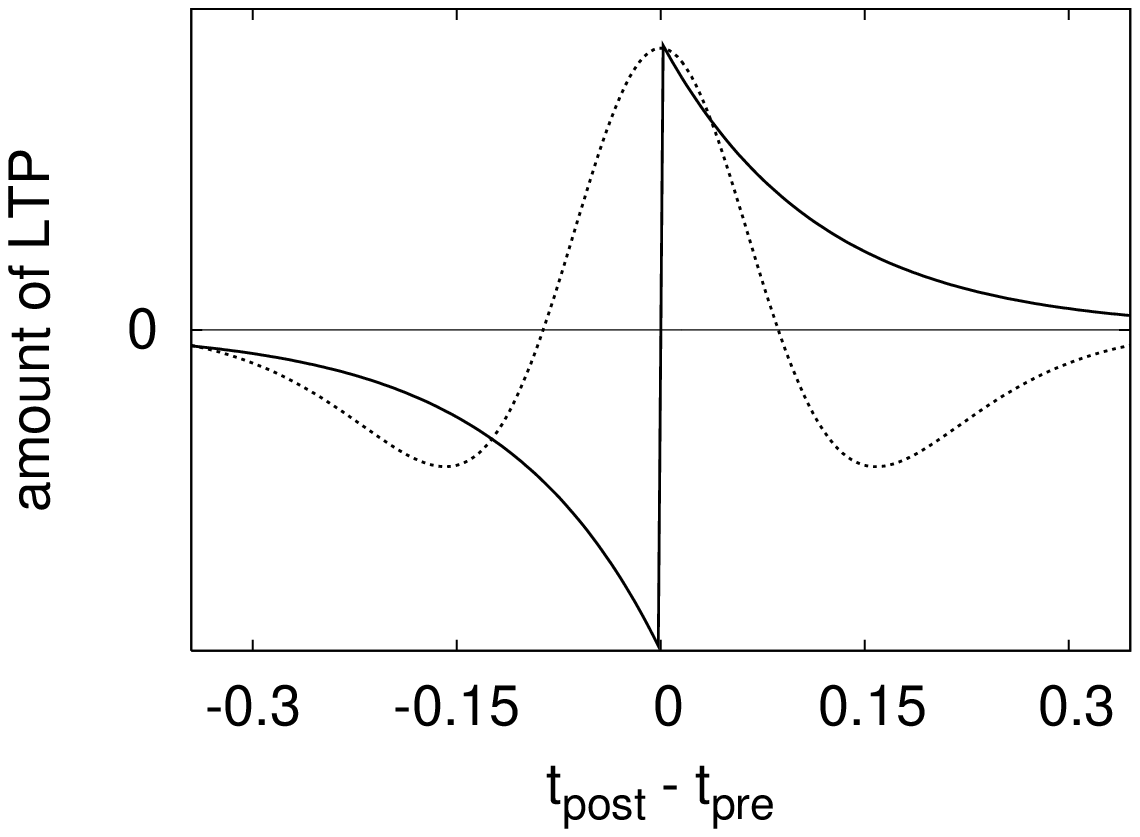}
\caption{Asymmetric (solid line) and symmetric (dashed line)
learning windows of STDP 
as a function of $t_{post}-t_{pre}$, namely, the postsynaptic spike
time relative to the presynaptic spike time.}
\label{fig:windows}
\end{center}
\end{figure}

Asymmetric STDP is modeled as follows.
LTP is induced if a presynaptic firing
(spike of oscillator $j$) precedes a postsynaptic firing (spike of
oscillator $i$).  In the opposite case, LTD occurs.  We denote the
presynaptic (postsynaptic) spike time by $t_{pre}$ ($t_{post}$).
A spike-pair event modifies the synaptic weight:
$g_{ji} \to g_{ji} + \Delta g_{ji}$, where
\begin{equation}
\Delta g_{ji} = 
\left\{ \begin{array}{ll}
A^+ \exp\left(-\frac{t_{post}-t_{pre}}{\tau}\right), &
t_{pre}<t_{post},\\
- A^- \exp\left(-\frac{t_{pre}-t_{post}}{\tau}\right), &
t_{pre}>t_{post},
\end{array}\right.
\label{eq:stdp}
\end{equation}
under the limitation $g_{ji}\in [0,g_{max}]$. A sample learning window is
indicated by the solid line in \FIG\ref{fig:windows}.
The width of the learning window is specified by
$\tau$, which
is known to be
of the order of 10--20 ms \cite{Bi98,Zhang98}. We confine ourselves
to the regime in which firing rates are not very large (5--20 Hz),
as is true for many pyramidal neurons.
Then, $\tau$
is several times smaller than the characteristic interspike
interval $T =$ 50--200 ms.
We thus set
\begin{equation}
\tau= \frac{1}{6}\times \frac{2\pi}{\Omega}\cong \frac{T}{6}.
\label{eq:set_tau}
\end{equation}
For completeness, we assume that $t_{pre} =
t_{post}$ does not induce plasticity. 

We assume that synaptic weights evolve so slowly that we can solve
\EQ(\ref{eq:homo-oscillators}) 
by regarding the synaptic weights as constant. This
assumption is valid if $A^+ \Omega, A^- \Omega \ll g^2$ for the
following reason. 
Because the relative phase relationship determines the evolution of
synaptic weights, we should compare the typical timescale of the
relative phase dynamics with that of synaptic plasticity.
The former is the
inverse of typical synaptic weight $g_0$. By introducing
dimensionless synaptic weight $g/g_0$, we find from \EQ(\ref{eq:stdp})
that the timescale of dimensionless synaptic plasticity is the
inverses of $A^+ \Omega/g_0$ and $A^- \Omega/g_0$.
The two timescales are
separated if $A^+ \Omega/g_0$, $A^- \Omega/g_0 \ll g_0$, which leads to
$A^+ \Omega$, $A^- \Omega \ll g_0^2 \cong g^2$. 
On the slow timescale of synaptic
plasticity, we can set $A^- = 1$ by rescaling the time, so that only the
ratio $A^+/A^-$ is relevant. For the stability,
$A^+/A^-$ must be balanced. This ratio is assumed to be 
slightly smaller than
unity according to previous literature \cite{Song00,Song01}.

Most of our theoretical efforts are invested in asymmetric STDP
because many pyramidal neurons show asymmetric STDP.
However, 
symmetric STDP, in which the synaptic update rule
depends only on $|t_{pre}-t_{post}|$, is also found
in some experiments.
Particularly, the learning
window is often shaped like a mexican hat in excitatory synapses; small
(large) $|t_{pre}-t_{post}|$ induces LTP (LTD)
\cite{Nishiyama,Abbott00,Shouval}. Symmetric learning windows have been
found for inhibitory synapses \cite{Woodin} and for the amount of LTD
in excitatory synapses \cite{Dan}.
We numerically 
analyze networks with symmetric STDP in \SEC\ref{sec:large}. 
We adopt superposition of two gaussian
distributions as the symmetric learning window, as depicted by the dotted
line in
\FIG\ref{fig:windows}. A spike-pair event modifies the synaptic
weight: $g_{ji} \to g_{ji} + \Delta g_{ji}$, where
\begin{equation}
\Delta g_{ji} = \frac{A^+}{\sqrt{2\pi\sigma^{+2}}}
\exp\left(-\frac{(t_{pre}-t_{post})^2}{2\sigma^{+2}}\right)
-
\frac{A^-}{\sqrt{2\pi\sigma^{-2}}}
\exp\left(-\frac{(t_{pre}-t_{post})^2}{2\sigma^{-2}}\right),
\label{eq:stdp-basic-rule-symm}
\end{equation}
with $\sigma^+ < \sigma^-$.  We set $\sigma^+= 0.6\tau$ and $\sigma^-
= 2 \sigma^+ = 1.2\tau$ so that the timescale of the symmetric
learning window is comparable to that of the asymmetric learning
window defined in \EQ(\ref{eq:stdp}).  The values of $A^+$ and $A^-$
are assumed to be the same as those for asymmetric STDP so that
$g_{ji}$ is bounded.

\section{Analysis of Small Networks with Asymmetric STDP}\label{sec:small}

We begin with small networks of two oscillators with
asymmetric STDP. For
these networks, how much initial coupling is necessary for
synchrony can be analytically evaluated.

\subsection{One Pacemaker and One Oscillator}\label{sub:p_and_o}

\begin{figure}
\begin{center}
\includegraphics[height=8.0cm,width=6.0cm]{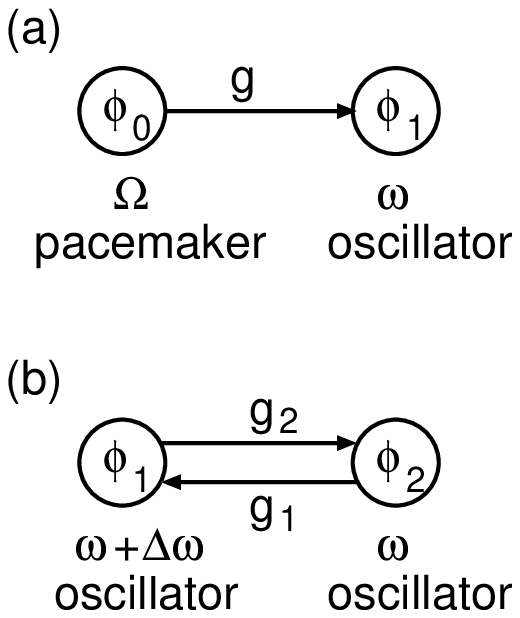}
\caption{Schematic diagrams showing
(a) the network of one pacemaker and one
oscillator, and (b) the network of two oscillators.}
\label{fig:diag_po_oo}
\end{center}
\end{figure}

We deal with the case $n=2$, namely, a network of one pacemaker and
one oscillator.  Because the connection from the oscillator to the
pacemaker does not affect the dynamics of the pacemaker, it suffices
to consider the unidirectional case. The network is schematically
shown in \FIG\ref{fig:diag_po_oo}(a).
We write $g=g_{01}$ to simplify
the notation.  The short-term dynamics in which $g$ is regarded to be
constant are described by
\begin{eqnarray}
\dot{\phi}_0 &=& \Omega,
\label{eq:a-p}\\
\dot{\phi}_1 &=& \omega + g \sin(\phi_0 - \phi_1).
\label{eq:a-1}
\end{eqnarray}
With $\psi \equiv \phi_0 - \phi_1$, \EQS(\ref{eq:a-p}) and
(\ref{eq:a-1}) reduce to
\begin{eqnarray}
\dot{\psi} &=& \Omega - \omega - g \sin\psi.
\label{eq:a-p1}
\end{eqnarray}
Based on the assumption that synaptic plasticity occurs much
more slowly than firing, we perform quasistatic analysis.
For a frozen $g$,
let us derive
the average angular frequency of the
oscillator denoted by $\tilde{\omega}$.
If 
$g\ge \Omega - \omega$, the pacemaker and the oscillator are in
frequency synchrony, i.e. 
$\tilde{\omega} = \Omega$.
If $0\le g <\Omega-\omega$, 
\EQ(\ref{eq:a-p1}) is equivalent to
\begin{equation}
\int \frac{d\psi}{\Omega - \omega - g \sin\psi} = \int dt.
\label{eq:dpsi-dt}
\end{equation}
Integration of \EQ(\ref{eq:dpsi-dt}) over a cycle yields
\begin{eqnarray}
\frac{2\pi}{\Omega-\tilde{\omega}} = \int^T_0 dt &=&  
\int^{2\pi}_{0} \frac{d\psi}{\Omega - \omega - g \sin\psi}\nonumber\\
&=& \frac{2\pi}{\sqrt{(\Omega-\omega)^2 - g^2}},
\end{eqnarray}
which results in
\begin{equation}
\tilde{\omega} = \Omega - \sqrt{ (\Omega-\omega)^2 - g^2}.
\label{eq:tildeomega_forward}
\end{equation}
Note that $\omega \le \tilde{\omega} <\Omega$.

The direction and the amount of synaptic plasticity induced by a single
spike-pair event is determined by $t_{post}-t_{pre}$.
We estimate $t_{post}-t_{pre}$ in terms of $\psi$ as follows.
Suppose that the phase difference is equal to $\psi$ when the
pacemaker fires. Then, it approximately takes $t_{post}-t_{pre}=
\psi/\tilde{\omega}$ for the oscillator to fire. In this case, LTP 
is induced because the pacemaker is presynaptic to the oscillator.
The pacemaker and the
oscillator can fire in the opposite order.
If the phase difference is $\psi$ when the
oscillator fires prior to the pacemaker does, the pacemaker
spends approximately
$t_{post}-t_{pre}=\psi/\Omega$ before firing. In this case,
LTD is induced. 
Because we confine ourselves to the case 
in which $\Omega$ does not deviate so
much from $\omega$, we approximate $t_{post}-t_{pre}\cong
\psi/\Omega$ regardless of the order of firing.

In \EQ(\ref{eq:set_tau}), we assumed
that the decay rate of the learning window
$\tau$ is
sufficiently smaller than $T/2$, which corresponds to
phase $\pi$. Therefore, the amount of
LTP is negligible for $t_{post}-t_{pre}\cong
\pi/\tilde{\omega}$ or larger, and the LTP rule is effective only
when $0<\psi\le\pi$. By the same token, the LTD
rule is effective only for $-\pi<\psi<0$.
Using these approximations, we aim to describe the dynamics of 
synaptic plasticity in terms of phase variables.
The amount of plasticity
given by \EQ(\ref{eq:stdp})
can be rewritten as
\begin{equation}
\Delta g = \left\{ \begin{array}{ll}
A^+ \exp\left(-\frac{\psi}{\Omega\tau}\right), &
0<\psi\le \pi, \; \phi_1 = 0\\
- A^- \exp\left(\frac{\psi}{\Omega\tau}\right), &
-\pi < \psi < 0, \; \phi_0 = 0
\end{array}\right.
\label{eq:rule-asym}
\end{equation}
where $\phi_1 = 0$ and $\phi_0=0$ indicate the postsynaptic spike time
for an LTP event (the spike time of the oscillator) and that of 
an LTD event (the spike time of the pacemaker), respectively.

We denote by $g(0)$ the initial synaptic weight.
If $g(0)\ge \Omega-\omega$,
fast dynamics have two steady states given by
\begin{equation}
\psi^* = \arcsin\left(\frac{\Omega-\omega}{g(0)} \right).
\end{equation}
The solution with  $\pi/2< \psi^*\le \pi$ is unstable, and hence the
fast dynamics converges to $\psi^*$ satisfying
$0 \le \psi^* \le \pi/2$. Therefore, STDP induces potentiation of $g$.
Then, $\psi^*$ for an altered $g$ becomes even smaller, which
induces further potentiation of $g$. Eventually, $g = g_{max}$ is
achieved. In sum, if
\begin{equation}
g(0)\ge g_c\equiv \Omega - \omega,
\end{equation}
the pacemaker and the oscillator will
synchronize quickly without plasticity.
The STDP does not break frequency synchrony.
Note that STDP generally decreases the phase difference $\psi$, but
$\psi$ does not tend to 0 (no 
phase synchrony) unless $\Omega=\omega$. Alternatively, 
if $g_{max} = \infty$,
$g$ diverges, and $\psi$ goes to 0.

Does asymmetric STDP facilitate frequency
synchrony?
If $g(0) <  g_c$, the oscillator is
initially not entrained by the pacemaker. Then,
$\psi$ slips.
By averaging over many spike-pair events, 
we represent the synaptic dynamics on a slow timescale by
\begin{eqnarray}
\dot{g} &=& \frac{\tilde{\omega}}{2\pi}\left[ \int_{t \; \; \mbox{
      such that } \; 0<\psi<\pi} A^+ 
e^{-\psi/\Omega\tau} dt
- \int_{t \; \; \mbox{ such that } \; -\pi< \psi < 0}
A^- e^{-\psi/\Omega\tau} dt\right]
\label{eq:a-dotgf-0}\\
&\propto& \int^{\pi}_0 e^{-\psi/\Omega\tau}
\left( \frac{A^+ }{g_c-g\sin\psi} 
- \frac{A^-}{g_c+g\sin\psi}\right) d\psi.
\label{eq:a-dotgf}
\end{eqnarray}
Derivation of \EQ(\ref{eq:a-dotgf-0}) requires the nonresonant 
situation. If $M_1 \Omega = M_2
\omega$ holds with small $M_1$ and $M_2$,
the dynamics become periodic with a rather small
period when the oscillator decouples from the pacemaker
due to STDP. If this were the case, the dependence on the initial
condition does not vanish permanently. In other words,
$\psi$ conditioned by $\phi_0=0$ or $\phi_1=0$
in \EQ(\ref{eq:rule-asym}) 
would take only limited values. Then the distribution of $\psi$
conditioned by a spike event would deviate from the unconditioned 
distribution of $\psi$.
With our choice of $\Omega = 9.1$ and
$\omega = 8.1$, the effect of such a resonance is very small.

In the region of $g$ where $\dot{g}>0$ holds, the RHS of
\EQ(\ref{eq:a-dotgf}) increases monotonically with $g$. If $g(0)$ is
greater than the value that makes the RHS equal to zero, which we
denote by $g_{c-stdp}$,
we obtain $\dot{g}>0$. Under this condition, $g$ continues to increase,
and $\psi^*$ decreases. This makes $\dot{g}$ in \EQ(\ref{eq:a-dotgf})
even larger. This positive feedback lasts until $g\ge g_c$ is
eventually satisfied. As a result, frequency synchrony is elicited by
STDP.  However, if $g(0)<g_{c-stdp}$, $g$ converges to the lower bound
0, so that the oscillator is disconnected from the pacemaker.

We bound $g_{c-stdp}$ as follows:
\begin{eqnarray}
&& \mbox{ RHS of \EQ(\ref{eq:a-dotgf}) }\nonumber\\
&=& \int^{\pi}_0 e^{-\psi/\Omega\tau}
\left[\frac{A^+}{g_c}\frac{1}{1-\frac{g\sin\psi}{g_c}}
- \frac{A^-}{g_c} \left(1-\frac{\frac{g\sin\psi}{g_c}}
{1+\frac{g\sin\psi}{g_c}} \right)
\right]\nonumber\\
&\ge& \int^{\pi}_0 e^{-\psi/\Omega\tau}
\left[ \frac{A^+}{g_c} \left(1+\frac{g\sin\psi}
{g_c}\right)
- \frac{A^-}{g_c}\left(1-\frac{g\sin\psi}{g_c}
\frac{1}{1+\frac{g}{g_c}}\right)
\right] d\psi\nonumber\\
&=& -\frac{A^- - A^+}{g_c}\int^{\pi}_0 e^{-\psi/\Omega\tau} d\psi
+\frac{g}{g_c^2}
\left( A^+ + \frac{A^-}{1+\frac{g}{g_c}}\right)
\int^{\pi}_0 e^{-\psi/\Omega\tau}\sin\psi d\psi
\nonumber\\
&=& - \frac{(A^- - A^+)\Omega\tau (1 - e^{-\pi/\Omega\tau})}{g_c}
+ \frac{g}{g_c^2} \frac{A^+ + A^-}{1+\frac{g}{g_c}}
\frac{\Omega^2\tau^2 (1 + e^{-\pi/\Omega\tau})}{1+\Omega^2\tau^2}.
\label{eq:RHS}
\end{eqnarray}
The value of $g$ that makes the RHS of the above equation zero gives
an upper bound of $g_{c-stdp}$.
When $A^+$ and $A^-$ are balanced
$(A^+ \cong A^-)$, we obtain
\begin{equation}
g_{c-stdp} \le
\frac{\frac{A^- - A^+}{A^- +
  A^+} \frac{(1+\Omega^2\tau^2)(1 - e^{-\pi/\Omega\tau})}
{\Omega\tau(1 + e^{-\pi/\Omega\tau})}} 
{1 - \frac{A^- - A^+}{A^- +
  A^+} \frac{(1+\Omega^2\tau^2)(1 - e^{-\pi/\Omega\tau})}
{\Omega\tau(1 + e^{-\pi/\Omega\tau})}}
g_c
\cong
\frac{A^- - A^+}{A^- +
  A^+} \frac{(1+\Omega^2\tau^2)(1 - e^{-\pi/\Omega\tau})}
{\Omega\tau(1 + e^{-\pi/\Omega\tau})} 
g_c.
\label{eq:gcstdp-linear}
\end{equation}  
Because $\Omega\tau$ is assumed to be of the order of $\pi$ (see
\EQ(\ref{eq:set_tau})), $\Omega\tau = O(1)$.
In addition, when $A^+ \cong A^-$, the inequalities in 
\EQS(\ref{eq:RHS}) and (\ref{eq:gcstdp-linear}) nearly hold with
the equalities. In such a case,
\begin{equation}
g_{c-stdp}\propto \left(1-\frac{A^+}{A^-}\right)g_c,
\end{equation}
which implies that $g_{c-stdp}$ is much smaller than $g_c$ when
$A^+\cong A^-$. Particularly, $g_{c-stdp}$ is extinguished when
$A^+\ge A^-$.  

\begin{figure}
\begin{center}
\includegraphics[height=6.0cm,width=6.0cm]{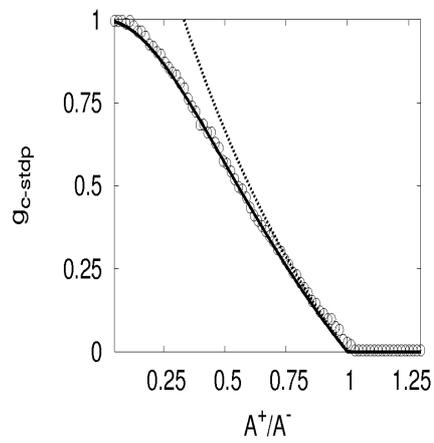}
\caption{$g_{c-stdp}$ for the network with
one pacemaker and one oscillator. The evaluation by
\EQ(\ref{eq:a-dotgf}) (solid line),
that by \EQ(\ref{eq:gcstdp-linear}) (dotted line), and
$g_{c-stdp}$ determined by numerical simulations of the coupled phase
oscillators (circles) are compared. We set
$g_{max} = 1.25$.}
\label{fig:p_and_o}
\end{center}
\end{figure}

In \FIG\ref{fig:p_and_o}, we plot $g_{c-stdp}$ evaluated by
numerical integration of \EQ(\ref{eq:a-dotgf}) and the
approximation given by \EQ(\ref{eq:gcstdp-linear}). We also plot
$g_{c-stdp}$ obtained by numerical simulations of our model
(\EQS(\ref{eq:stdp}),
(\ref{eq:a-p}), and (\ref{eq:a-1})), in which
$g_{c-stdp}$ is determined by varying
the initial synaptic weight $g(0)$.
The evaluation by \EQ(\ref{eq:a-dotgf})
(solid line) is in good agreement with $g_{c-stdp}$ obtained by 
numerical simulations of the model (circles)
for a broad range of
$A^+/A^-$. As expected, the approximate estimation by
\EQ(\ref{eq:gcstdp-linear}) (dotted line)
also agrees with the numerical data (circles)
when $A^+/A^-$ is close to unity. 
In conclusion, asymmetric
STDP drastically enhances frequency synchrony.

Regarding symmetric STDP, for values of $g$ such that $\psi$ falls in
the positive learning window (refer to the dotted line in
\FIG\ref{fig:windows}), $g$ is strengthened to eventually exceed
$g_c$.  Therefore, frequency synchrony is facilitated. To what extent
synchrony is promoted depends on the width of the learning window.

When $\Omega < \omega$, there are two solutions $\psi^* \in
(-\pi,0)$, one of which is stable.  Then, STDP elicits
LTD. Even though $\psi^*$ changes, the relation $-\pi< \psi^* < 0$ is
preserved until the pacemaker and the oscillator get disconnected.  As
a result, frequency synchrony does not happen.

\subsection{Two Oscillators}\label{sub:o_and_o}

To examine how the connectivity between a pair of oscillators
evolves in a large network, we analyze the following toy model of two
bidirectionally coupled oscillators:
\begin{eqnarray}
\dot{\phi}_1 &=& \omega + \Delta\omega 
+ g_1\sin(\phi_2 - \phi_1),\nonumber\\
\dot{\phi}_2 &=& \omega + g_2\sin(\phi_1 - \phi_2), 
\label{eq:two-oscillators}
\end{eqnarray}
where $g_1$, $g_2$ $\in [0,g_{max}]$.  The network is depicted in
\FIG\ref{fig:diag_po_oo}(b).  Now two oscillators influence each
other, which contrasts to the case of the pacemaker-oscillator network
examined in \SEC\ref{sub:p_and_o}.  The term $\Delta \omega$
represents the mismatch in natural frequencies.  Although the
oscillators are identical in our original model (see
\EQ(\ref{eq:homo-oscillators})), we introduce $\Delta \omega$ because
of the following reason. In oscillator networks with a pacemaker, the
oscillators are not completely phase synchronized.  The oscillators
directly connected to the pacemaker are the first to fire after the
pacemaker does. Then, other oscillators adjacent to those connected to
the pacemaker fire after some delay, and so
forth. Therefore, the oscillators closer to the pacemaker tend to have
more advanced phases, and the distribution of the phases is associated with
the hierarchical organization of the network. Imagine two oscillators
coupled unidirectionally or bidirectionally in a large network. We
denote one that fires first and the other by oscillators 1 and 2
respectively. Precisely, the difference in the firing timing stems
from complex effects of coupling with other oscillators. For
analytical tractability, here we replace such effects by the frequency
mismatch $\Delta \omega$, by which the difference in the firing timing
can be easily introduced. As shown in the following, for $\Delta
\omega>0$, oscillator 1 tends to fire in advance of oscillator
2. Accordingly, we regard that oscillator 1 is closer to the pacemaker
than oscillator 2.

We analyze the model given by \EQ(\ref{eq:two-oscillators}).
By introducing $\psi \equiv \phi_1 - \phi_2$, we derive
\begin{equation}
\dot{\psi} = \Delta\omega - (g_1+g_2)\sin\psi.
\end{equation}

If
\begin{equation}
g_1(0) + g_2(0) \ge \Delta\omega,
\label{eq:2o-init-syn}
\end{equation}
two oscillators are
locked with phase lag $0< \psi^* <\pi/2$, where
\begin{equation}
\psi^* = \arcsin\left(\frac{\Delta\omega}{g_1 + g_2}\right).
\end{equation}
If synaptic plasticity is absent, \EQ(\ref{eq:2o-init-syn}) gives the
condition for frequency synchrony.

In contrast to the network of one pacemaker and one oscillator
analyzed in \SEC\ref{sub:p_and_o},
\EQ(\ref{eq:2o-init-syn})
does not guarantee that frequency synchrony is maintained
throughout STDP. When \EQ(\ref{eq:2o-init-syn}) is satisfied,
the synaptic dynamics are written as 
\begin{eqnarray}
\dot{g}_1 &=& -A^- \exp\left(-\frac{\psi^*}{\tilde{\omega}\tau}\right),\\
\dot{g}_2 &=& A^+ \exp \left(-\frac{\psi^*}{\tilde{\omega}\tau}\right),
\end{eqnarray}
where 
\begin{equation}
\tilde{\omega} = \omega + \frac{g_2\Delta\omega}{g_1+g_2}
\label{eq:tilde-omega-2o}
\end{equation}
is the frequency common to the two oscillators.
Because $A^+ < A^-$,
$g_1 + g_2$ decreases with time.
The oscillators desynchronize in frequency if $g_1+g_2\ge \Delta\omega$
is violated via synaptic plasticity.
For sufficiently small $g_1(0) +
g_2(0)$, the two oscillators are disconnected
even from the beginning. In these cases,
$\psi$ slips due to the absence of
frequency synchrony.

The average frequencies of the two oscillators out of frequency synchrony 
are calculated as
\begin{eqnarray}
\tilde{\omega}_1 &=& \omega + 
\frac{g_2\Delta\omega + g_1
\sqrt{\Delta\omega^2-(g_1+g_2)^2}}{g_1+g_2},\\
\tilde{\omega}_2 &=& \omega +
\frac{g_2\Delta\omega - g_2
\sqrt{\Delta\omega^2-(g_1+g_2)^2}}{g_1+g_2}.
\end{eqnarray}
The synaptic weights evolve according to
\begin{eqnarray}
\dot{g}_1 &=& \frac{1}{2\pi}
\int^{\pi}_{0} 
\left[ \frac{A^+ e^{-\psi/\tilde{\omega}_1\tau}}
{\Delta\omega+(g_1+g_2)\sin\psi}
- \frac{A^- e^{-\psi/\tilde{\omega}_2\tau}}
{\Delta\omega-(g_1+g_2)\sin\psi}
\right] d\psi\nonumber\\
&=& 
\frac{1}{2\pi}
\int^{\pi}_{0} 
e^{-\psi/\tilde{\omega}_1\tau}
\left[ \frac{A^+}{\Delta\omega+(g_1+g_2)\sin\psi}
- \frac{A^-}{\Delta\omega-(g_1+g_2)\sin\psi}
\right] d\psi,\\
\dot{g}_2 &=&
 \frac{1}{2\pi}
\int^{\pi}_{0}
\left[ \frac{A^+ e^{-\psi/\tilde{\omega}_2\tau}}
{\Delta\omega-(g_1+g_2)\sin\psi}
- \frac{A^- e^{-\psi/\tilde{\omega}_1\tau}}
{\Delta\omega+(g_1+g_2)\sin\psi}
\right] d\psi\nonumber\\
&=& \frac{1}{2\pi}
\int^{\pi}_{0}
e^{-\psi/\tilde{\omega}_1\tau}
\left[ \frac{A^+}{\Delta\omega-(g_1+g_2)\sin\psi}
- \frac{A^-}{\Delta\omega+(g_1+g_2)\sin\psi}
\right] d\psi,
\end{eqnarray}
where we approximated $e^{-\psi/\tilde{\omega}_2\tau}$ 
by $e^{-\psi/\tilde{\omega}_1\tau}$, as we did
in \SEC\ref{sub:p_and_o}.

Since $\dot{g}_1<0$ is always satisfied, $g_1$ eventually reaches 0;
backward connectivity from a downstream oscillator to an upstream
oscillator is eliminated. Whether a `forward' connectivity from
the upstream oscillator to the downstream oscillator survives relies
on $g_2(\hat{t})$ where $\hat{t}$ is the time $g_1$ reaches 0.
If $g_2(\hat{t})$ is larger than $g_{c-stdp}$ obtained in
\SEC\ref{sub:p_and_o}, frequency synchrony will be eventually
established.  In this case, the final oscillation frequency is equal
to $\omega + \Delta\omega$ so that oscillator 2 is enslaved by
oscillator 1.  If $g_2(\hat{t})<g_{c-stdp}$, the two oscillators are
finally disconnected.

\begin{figure}
\begin{center}
\includegraphics[height=6.0cm,width=6.0cm]{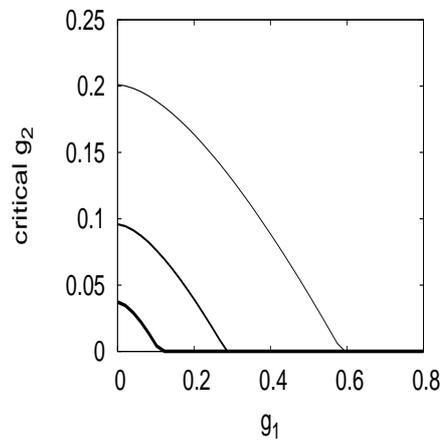}
\caption{The critical $g_2(0)$ as a function of $g_1(0)$ for 
the network with two oscillators (and no pacemaker).
Three lines correspond to 
$A^+/A^- = 0.96$ (thickest line), $0.9$,
and $0.8$ (thinnest line).}
\label{fig:o_and_o}
\end{center}
\end{figure}

The critical value of $g_2(0)$ above which frequency synchrony occurs is
plotted in \FIG\ref{fig:o_and_o} for different values of $A^+/A^-$
and $g_1(0)$.
The critical $g_2(0)$ decreases with $g_1(0)$, implying that
a large $g_1(0)$ enhances frequency synchrony.
Such backward connectivity
transiently serves to keep $\psi$ small so that
forward connectivity is enhanced.
However, only the synapse from the faster
oscillator to the slower oscillator survives eventually.

The feedforward network is not created via symmetric
STDP, by which $g_1$ and $g_2$ evolve in the same direction.
When initial 
mutual connectivity is strong enough,
synchrony is established 
so that the two synaptic weights are saturated
($g_1 = g_2 = g_{max}$). Then, based on \EQ(\ref{eq:tilde-omega-2o}),
the common firing frequency is equal to $\omega + \Delta\omega/2$,
but not to 
the frequency of the faster oscillator ($=\omega + \Delta\omega$).

The effect of LTP-LTD balance
is also shown in \FIG\ref{fig:o_and_o}. When $A^+$ is close to $A^-$,
critical $g_2(0)$ is lowered, and entrainment occurs easily.
Even when LTP is rather weak compared to LTD
($A^+/A^- = 0.8$, thinnest line), the critical 
$g_2$ is much reduced from $g_c$.

\section{Numerical Results for Large Networks}\label{sec:large}

So far, we have analyzed small networks composed of two elements only.
In this section, we examine how frequency synchrony can be
facilitated by STDP in larger networks. In particular, we compare
$g_c$ and $g_{c-stdp}$ and also investigate evolution of network
structure. To this end, we numerically simulate 
randomly
connected $n=100$ elements (99 oscillators and one pacemaker)
based on \EQS(\ref{eq:stdp}),
(\ref{eq:a-p}), and (\ref{eq:a-1}).

\subsection{Initial Setup}

We generate a directed random network as follows. Starting from a set
of $n$ isolated vertices, we add a directed edge that connects a randomly
chosen pair of oscillators. We forbid multiple directed edges between
a pair of oscillators and self loops, i.e. edges whose origin and
destination are identical.  This procedure is repeated
$n\left<k\right>$ times. In the following, we assume that
$\left<k\right>= 10$.  In other words, each
oscillator is presynaptic to 10 other oscillators and postsynaptic to
the same number of oscillators on average.

We define the distance
$l_i$ of oscillator $i$ from the pacemaker by the smallest number of
directed edges necessary to reach from the pacemaker to oscillator
$i$. For example, the number of the oscillators at distance 1 is equal
to those that receive direct synaptic contacts from the
pacemaker. Therefore, about $\left<k\right>$ oscillators have
distance 1. Among the other oscillators, those receiving
an edge from an oscillator
with distance 1 have distance 2. 
The depth $L$ of a network is defined as the
distance averaged over all the oscillators:
$L=\sum^{n-1}_{i=1} l_i /(n-1)$.

The initial phases $\phi_i$ ($0\le i\le n-1$) are
taken independently from the uniform density on $[0,2\pi)$.
For all the synapses, we initially set $g_{ji} = g(0)$.

For a specific random network used in the following numerical
simulations, we obtained $L=3.22$. For this network,
we numerically found that frequency synchrony happens without synaptic
plasticity if $g\ge 
g_c\cong 100.7$. In this case, the
pacemaker first fires in each cycle, and oscillators with smaller
distances tend to fire with smaller lag with respect to the
pacemaker \cite{Kori04}.

\subsection{Measured Quantities}\label{sub:order_parameters}

We define the degree of frequency synchrony $r \equiv
r\left([t_1, t_2]\right)$ 
for a time interval $[t_1, t_2]$.
The mean frequency of each oscillator for this time interval
is equal to
\begin{equation}
\tilde{\omega}_i = \frac{\phi_i(t_2)-\phi_i(t_1)}{t_2-t_1}.
\end{equation}
We note that $\tilde{\omega}_0 = \Omega$. Then, the synchrony measure
is defined by 
\begin{equation}
r = \frac{ \frac{1}{n-1}\sum^{n-1}_{i=1}
\tilde{\omega}_i - \omega}{\Omega-\omega}.
\end{equation}
When the oscillators are in frequency synchrony with the pacemaker,
the mean frequency of the oscillators
$\sum^{n-1}_{i=1} \tilde{\omega}_i / (n-1)$ is equal to the
frequency of the pacemaker $\Omega$, and we have $r = 1$.  If the
pacemaker does not at all affect the other $n-1$ oscillators, the
oscillators fire at their natural frequency $\omega$, and we have
$r=0$.  We divide the total simulation time into consecutive bins of
the width $t_2-t_1 = 100$ for oscillator simulations in
\SECS\ref{sub:asym} and \ref{sub:sym}.

More microscopically, we inspect possible formation of feedforward
chains originating from the pacemaker. To quantify this, we track
several order parameters derived from synaptic weights.  The first is
the depth $L$ extended to networks with heterogeneous
synaptic weights in the following way.  Let us consider a path from
the pacemaker to oscillator $i$.  A path is equivalent to a chain of
directed synapses: $(j_0,j_1)\in E$, $(j_1, j_2)\in E$, $\ldots$,
$(j_{k_i-1}, j_{k_i})\in E$, where $j_0=0$ and
$j_{k_i}=i$.  The length of this path is given by
$\sum^{k_i-1}_{k=0} g_{max}/g_{j_k j_{k+1}}$.  The distance $l_i$
of oscillator $i$ from the pacemaker is the shortest path length among
all possible paths from the pacemaker to oscillator $i$
\cite{Braunstein03}.
This
definition generalizes the prior definition for networks
with unit synaptic weights. The redefined distance is associated with
how much a downstream oscillator is influenced by the pacemaker. 
The depth of the
network is again defined by $L=\sum^{n-1}_{i=1} l_i / (n-1)$
and measures effective proximity of the oscillators from the
pacemaker.  By definition, the generalized $L$ is equal to or larger
than $L$ of the unweighted network, with equality realized only when
$g_{ji}=g_{max}$ for all the synapses that appear in the shortest paths.

A synaptic 
connection $(j,i)$ with $l_j<l_i$ ($l_j>l_i$) is a forward (backward)
connection in the meaning that it complies with the feedforward chain
emanating from the pacemaker. Accordingly, we define the amount of
forward connection $w_f$, that of backward connection $w_b$, and
that of lateral connection $w_l$ by
\begin{eqnarray}
G_f &=& \sum_{l_i-l_j>\epsilon} \frac{g_{ji}}
{n\left<k\right>},\label{eq:G_f}\\
G_b &=& \sum_{l_i-l_j<-\epsilon} \frac{g_{ji}}
{n\left<k\right>},\label{eq:G_b}\\
G_l &=& \sum_{-\epsilon\le l_i-l_j\le \epsilon} \frac{g_{ji}}
{n\left<k\right>}.\label{eq:G_p}
\end{eqnarray}
Summation is taken over the pairs of oscillators forming synapses
($(j,i)\in E$). Note that $G_l$ quantifies the connection between
oscillators whose distances from the pacemaker are approximately equal. 
 The number of synapses in the network
($=n\left<k\right>$) gives normalization, and thus $0\le G_f, G_b,
G_p\le g_{max}$. 
The average synaptic weight is given by $0\le G_f+G_b+G_p\le
g_{max}$.
The tolerance level is chosen to be sufficiently
small: $\epsilon=0.05$. 
We also define local quantities to evaluate feedforwardness.
The average weight of the synapses postsynaptic to the pacemaker is
denoted by $G_f^0$. This is equal to the average of $g_{0i}$ over
$i$ with $(0,i)\in E$. This corresponds to $g$ used in
\SEC\ref{sub:p_and_o}. Similarly, the average weight
of the synapses presynaptic to the pacemaker is denoted by
$G_b^0$. This is equal to the average of $g_{i0}$ over $i$ with 
$(i,0)\in E$. We note that $0\le G_f^0, G_b^0 \le g_{max}$.

\subsection{Asymmetric Learning Window}\label{sub:asym}

We apply asymmetric STDP with LTD slightly stronger than
LTP \cite{Song00,Song01}:
$A^+ = 0.009$ and $A^- = 0.01$.
By setting $g_{max} = 15 < g_c$, homogeneous enhancement of all the
synapses does not lead to
synchrony. We determine
$g_{c-stdp}$ by running numerical simulations with various values of 
the initial synaptic weight $g_{ji} = g(0)$.

Because $A^+$ is close to $A^-$, our results in \SEC\ref{sec:small}
predict the following.

\begin{itemize}

\item As shown in
\SEC\ref{sub:o_and_o}, backward connection will be eliminated via the
asymmetric STDP so that $G_b$ and
$G_b^0$ decrease.

\item The unidirectional
connection between the pacemaker and the
oscillator will be easily established (\SEC\ref{sub:p_and_o}). As a result,
a feedforward chain rooted at the pacemaker is expected to form.

\item $g_{c-stdp}$ is much smaller than $g_c$.

\end{itemize}

\begin{figure}
\begin{center}
\includegraphics[height=6.0cm,width=6.0cm]{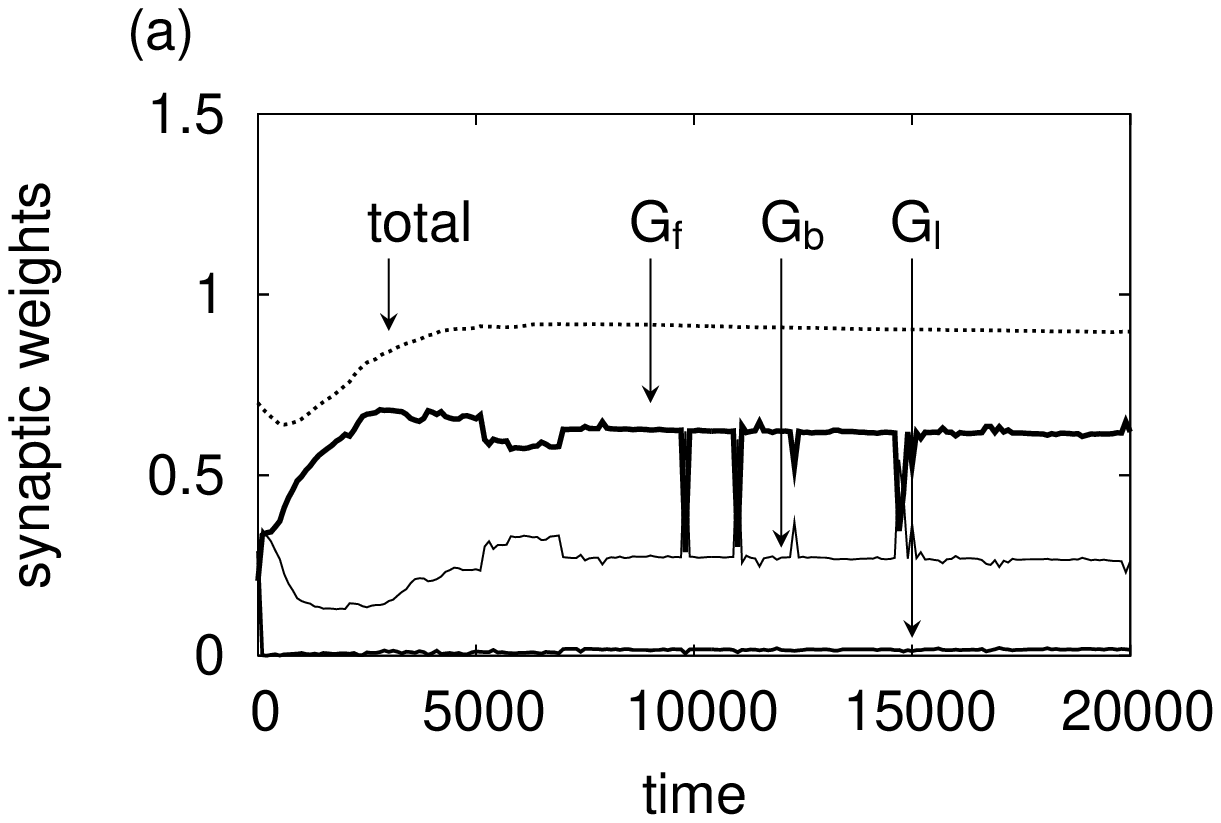}
\includegraphics[height=6.0cm,width=6.0cm]{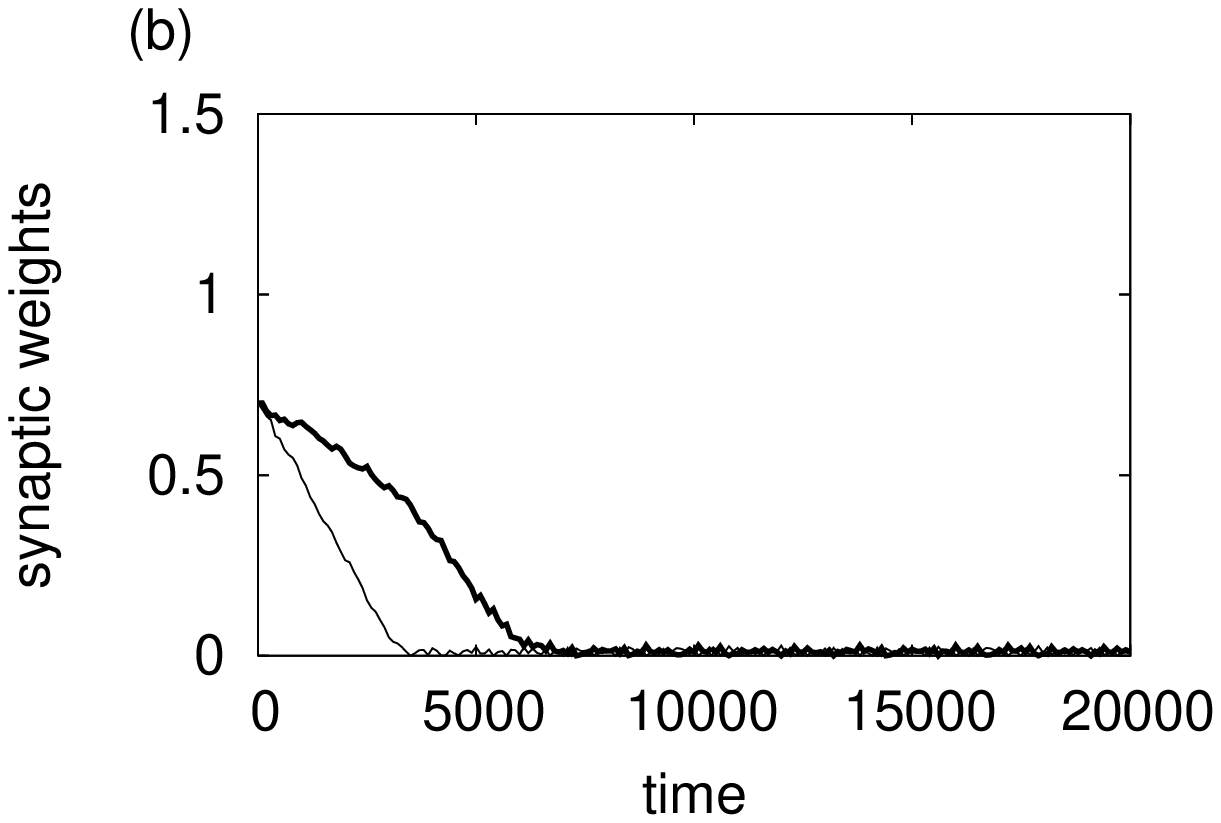}
\includegraphics[height=6.0cm,width=6.0cm]{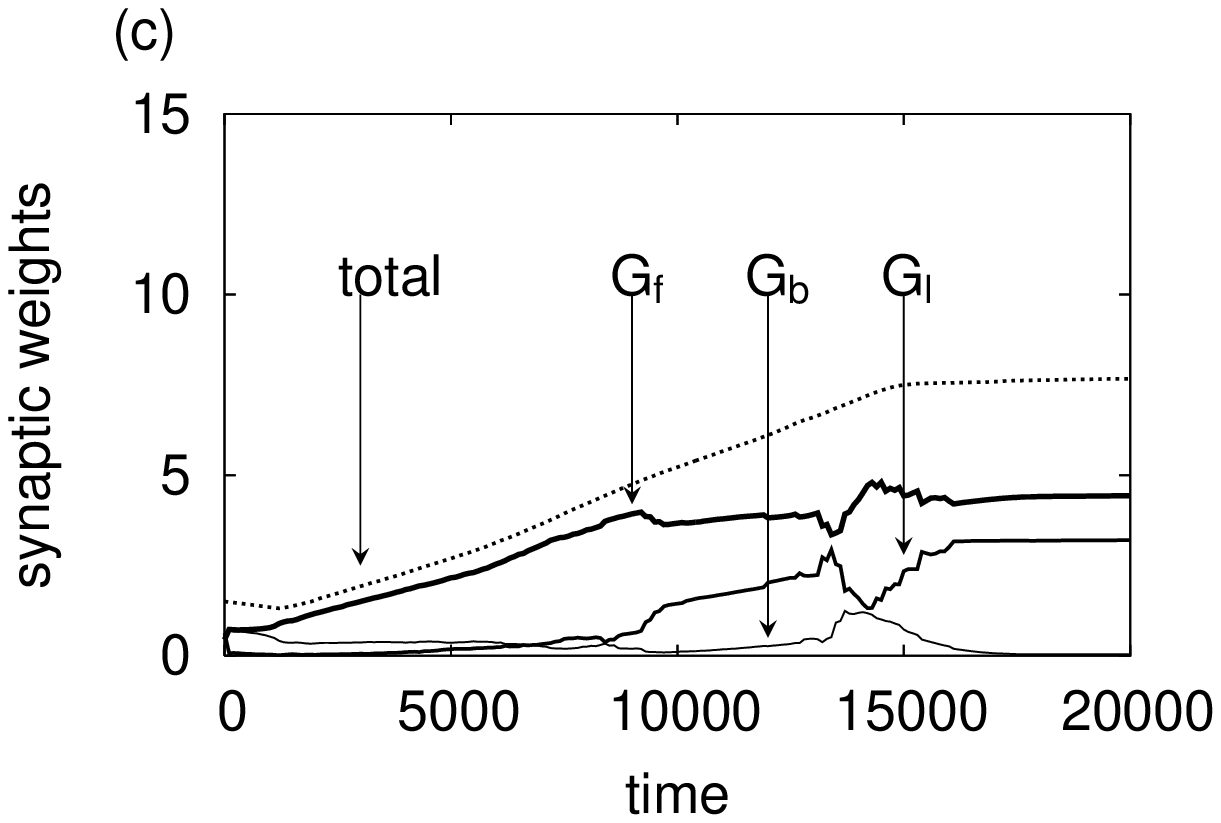}
\includegraphics[height=6.0cm,width=6.0cm]{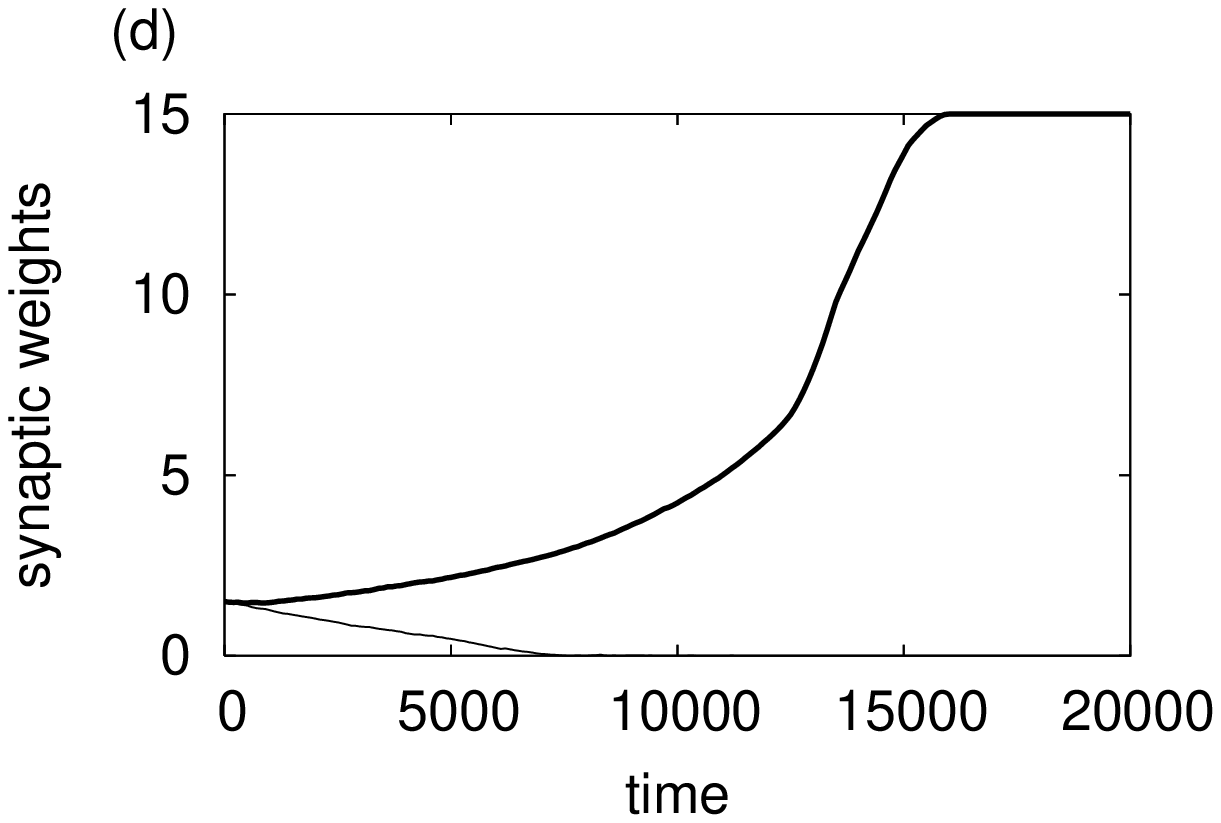}
\caption{Results for 100 randomly coupled oscillators subject to
asymmetric STDP. Evolution of the synaptic-weight order parameters are
shown for (a, b) $g(0)=0.7$ and (c, d) $g(0)=1.5$.  In (a) and (c),
$G_f$ (thick solid lines), $G_b$ (thin solid lines), $G_l$ (moderate
solid lines), and the average weight (dotted lines) are shown.  In
(b) and (d), $G_f^0$ (thick lines) and $G_b^0$ (thin
lines) are indicated.  Time courses of (e) $L$ and (f) $r$ are compared
for $g(0)=0.7$, 0.9, 1, 1.2, and 1.5.
In (e), lower lines correspond to larger $g(0)$.
 (g) Final network structure for $g(0)=1.5$.
Only the synapses $(j,i)\in E$ with
$g_{ji}>g(0)$ are presented. The pacemaker
is labeled $P$. Forward edges and backward edges are indicated
by thin lines and thick lines, respectively.
The network is drawn by Pajek
(http://vlado.fmf.uni-lj.si/pub/networks/pajek/).
(h) Time courses of $r$ for some values of $g(0)$
when the oscillators are heterogeneous.}
\label{fig:asym}
\end{center}
\end{figure}

Dynamics of synaptic-weight order parameters for $g(0)=0.7$ are shown
in \FIG\ref{fig:asym}(a, b).  The average synaptic weight
(dotted line in \FIG\ref{fig:asym}(a)) increases in the
initial stage because some synapses between the oscillators are
potentiated.  However, its stationary value is much smaller than
the maximal possible value ($g_{max}=15$).
There is no selective potentiation of forward synapses 
($G_f$, thick solid line in \FIG\ref{fig:asym}(a))
or depression of backward synapses ($G_b$, thin solid line in
\FIG\ref{fig:asym}(a)). The forward
connectivity from the pacemaker
($G_f^0$, thick line in \FIG\ref{fig:asym}(b)) also
degrades with time.  Eventually, the
oscillators disconnect from the pacemaker, which is
observed as indefinitely growing $L$ (uppermost line in
\FIG\ref{fig:asym}(e)).  Accordingly, frequency synchrony between the
pacemaker and the oscillators is not achieved; \FIG\ref{fig:asym}(f)
indicates that $r$ stays near $0$.

\begin{figure}
\begin{center}
\includegraphics[height=6.0cm,width=6.0cm]{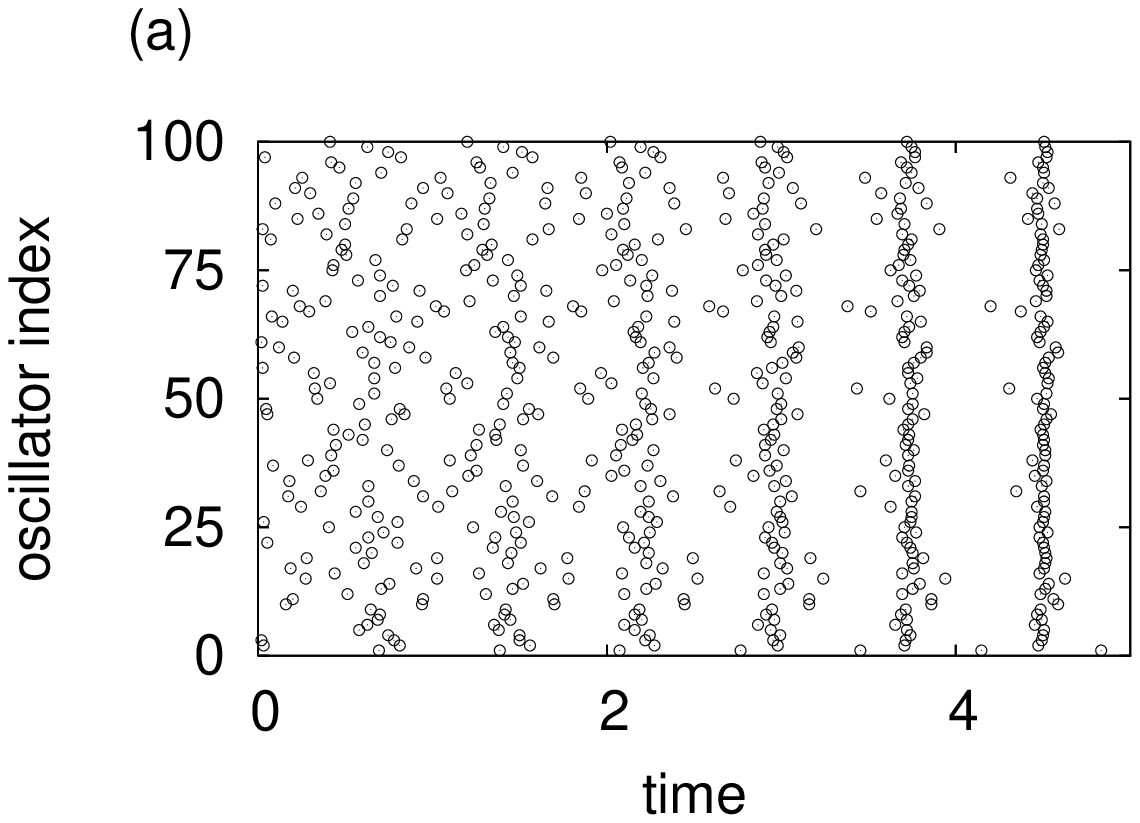}
\includegraphics[height=6.0cm,width=6.0cm]{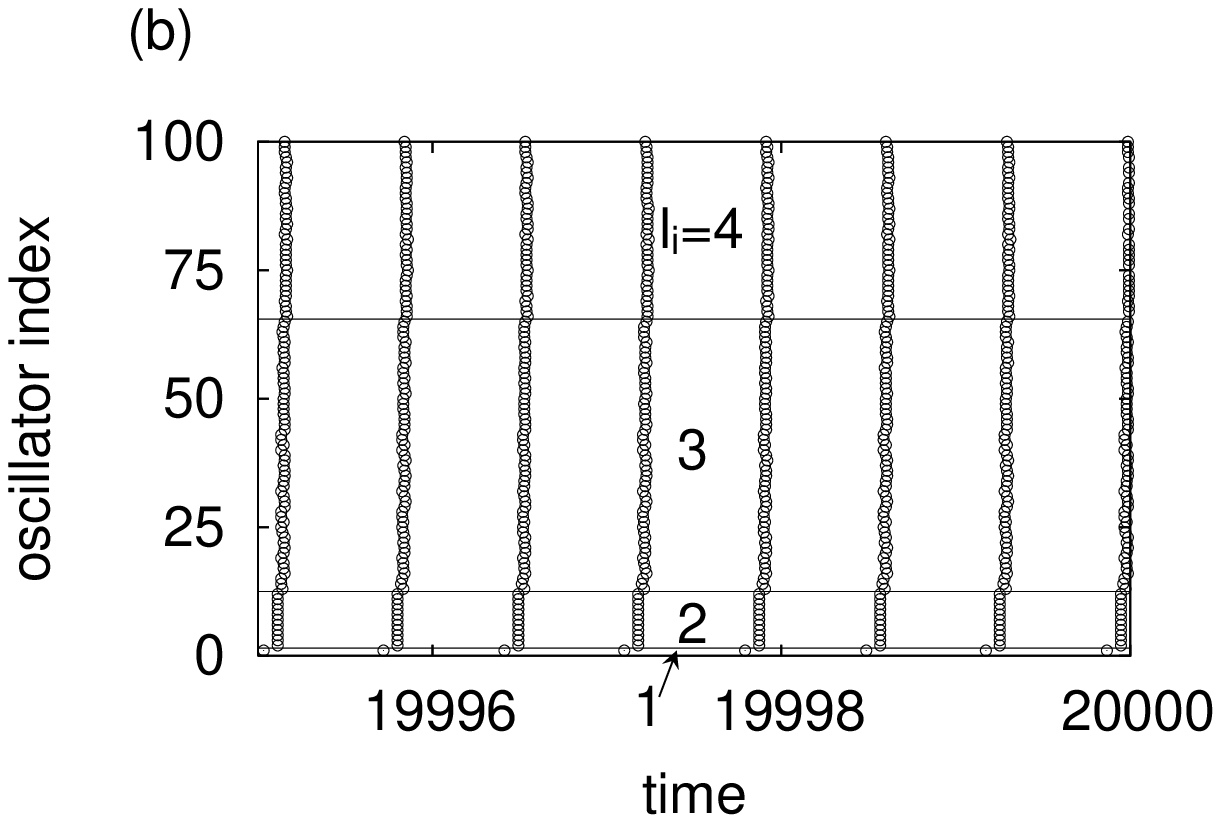}
\includegraphics[height=6.0cm,width=6.0cm]{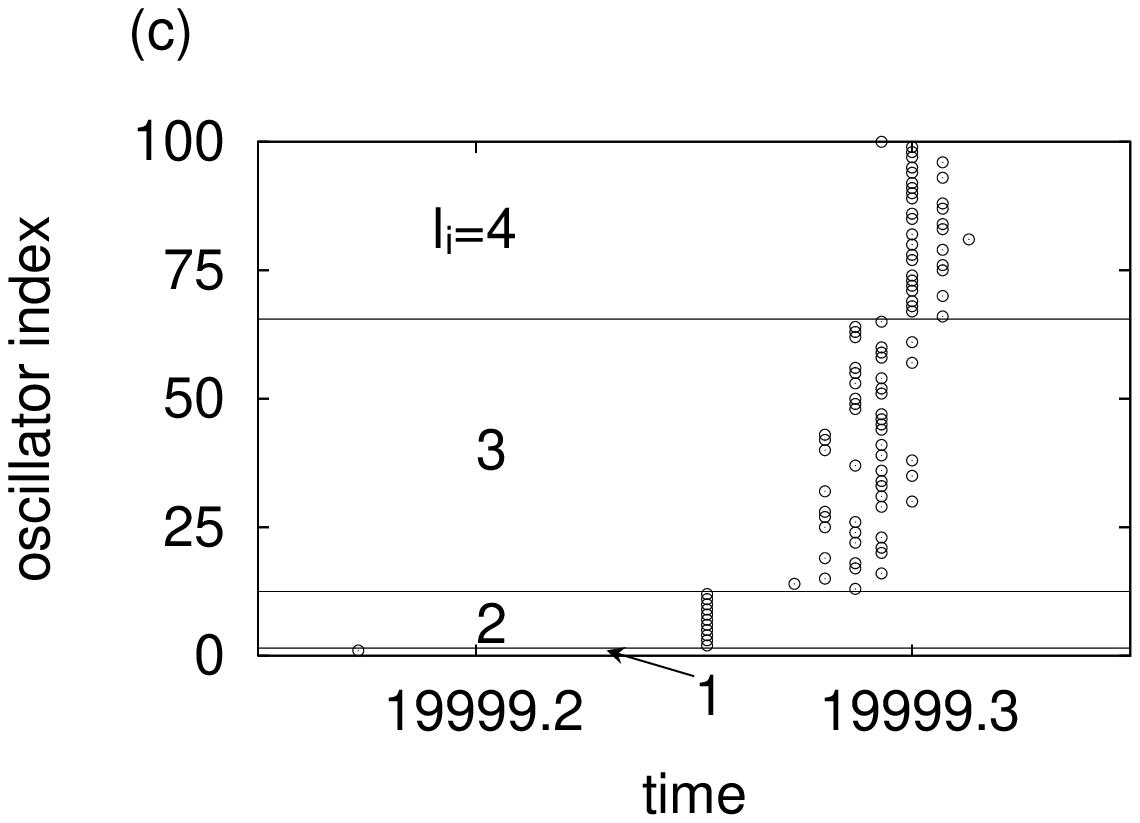}
\caption{Rastergrams of the oscillators under asymmetric STDP
in (a) initial and (b) final cycles. We set $g(0)=1.5$.
The oscillators 
are aligned 
according to their distances $l_i$ from the pacemaker, which is
calculated at time 0 in (a) and 19995 in (b).
After sufficient time,
$l_i$ is quantized, and the values of $l_i$ are shown in (b). (c)
is a magnification of (b).}
\label{fig:raster}
\end{center}
\end{figure}

By contrast, frequency synchrony without phase synchrony is 
established when $g(0) = 1.5$, 
as supported by the rastergrams in \FIGS\ref{fig:raster}(a)
and \ref{fig:raster}(b) corresponding to
initial and final periods of simulations,
respectively. More in detail,
forward connectivity $G_f$ (thick
solid line in \FIG\ref{fig:asym}(c)) and $G_f^0$ (thick line in
\FIG\ref{fig:asym}(d)) grow toward $g_{max}$ to result in frequency
synchrony at $t\cong 12500$ (\FIG\ref{fig:asym}(f)), 
accompanied by a decrease in $L$ (lowermost
line in \FIG\ref{fig:asym}(e)). Backward synapses directly projecting to
the pacemaker are pruned in an
initial stage ($G^0_b$; thin line in \FIG\ref{fig:asym}(d)). It takes
longer time for $G_b$ to decay (thin solid line in
\FIG\ref{fig:asym}(c)).  Although randomness
in the initial condition blurs the phase transition, we estimate
$g_{c-stdp}\cong 0.9$ based on \FIGS\ref{fig:asym}(e) and
(f). Consistent with the results in \SEC\ref{sub:p_and_o},
$g_{c-stdp}$ is much smaller than $g_c \cong 100.7$.

Frequency synchrony is made possible by
combined effects of sufficiently large forward weights and
sufficiently small backward weights.  It is not an immediate
consequence of increased average synaptic weights; achieving synchrony
merely by homogeneously strong synapses necessitates $G_f + G_b + G_p
\ge g_c$. Because of our choice of $g_{max}$ ($<g_c$),
homogeneous LTP does not induce frequency synchrony.  Elimination of
backward weights is essential for frequency synchrony.  The final
network structure reconstructed from synapses with $g_{ji}>g(0) = 1.5$ is
shown in \FIG\ref{fig:asym}(g), with forward and backward edges shown
by thin and thick lines, respectively.  Few backward edges survive
asymmetric STDP.  The network is close to a feedforward network rooted
at the pacemaker, which enslaves the oscillators.  

We remark that detailed behavior of the network order
parameters varies according to the value of 
$\epsilon$ and the definition of the distance
$l_i$, which is inherently arbitrary for weighted networks. However,
the general tendency that forward synapses are potentiated and
backward synapses are depressed for $g\ge g_{c-stdp}$ is observed
irrespective of these details. For a reasonably defined
$l_i$, whether $L$ increases or decreases
(\FIG\ref{fig:asym}(e)) and whether
feedforward networks form (\FIG\ref{fig:asym}(g)) are determined
independently of the definition of $l_i$. 

We have performed additional 
numerical simulations in which every presynaptic
spike spends for $\tau/5 = T/30$ before exciting the postsynaptic neurons.
Because $T =$ 50--200 ms, the corresponding synaptic delay
is equal to a few milliseconds.
The value of $g_{c-stdp}$ 
hardly changes with this synaptic delay (results not shown).

To examine the effect of heterogeneity in oscillators, we pick the
intrinsic frequency of each oscillator from the gaussian distribution
with mean $\omega = 8.1$ and standard deviation 0.1. Time courses of
$r$ are shown in \FIG\ref{fig:asym}(h).
We estimate $1.2<g_{c-stdp}<1.5$,
implying the robustness of our results against heterogeneity.  We remark
that $r$ converges to a positive level
when $g<g_{c-stdp}$. This is because, even
if the oscillators disconnect from the pacemaker, some
oscillators form feedforward networks of small size in which fast
oscillators entrain and speed up slow oscillators.  If the
heterogeneity is even larger so that some oscillators are as fast as
the pacemaker, frequency synchrony seeding from the pacemaker
would be difficult because the
pacemaker and fast oscillators compete in entraining slow oscillators.

\subsection{Symmetric Learning Window}\label{sub:sym}

\begin{figure}
\begin{center}
\includegraphics[height=6.0cm,width=6.0cm]{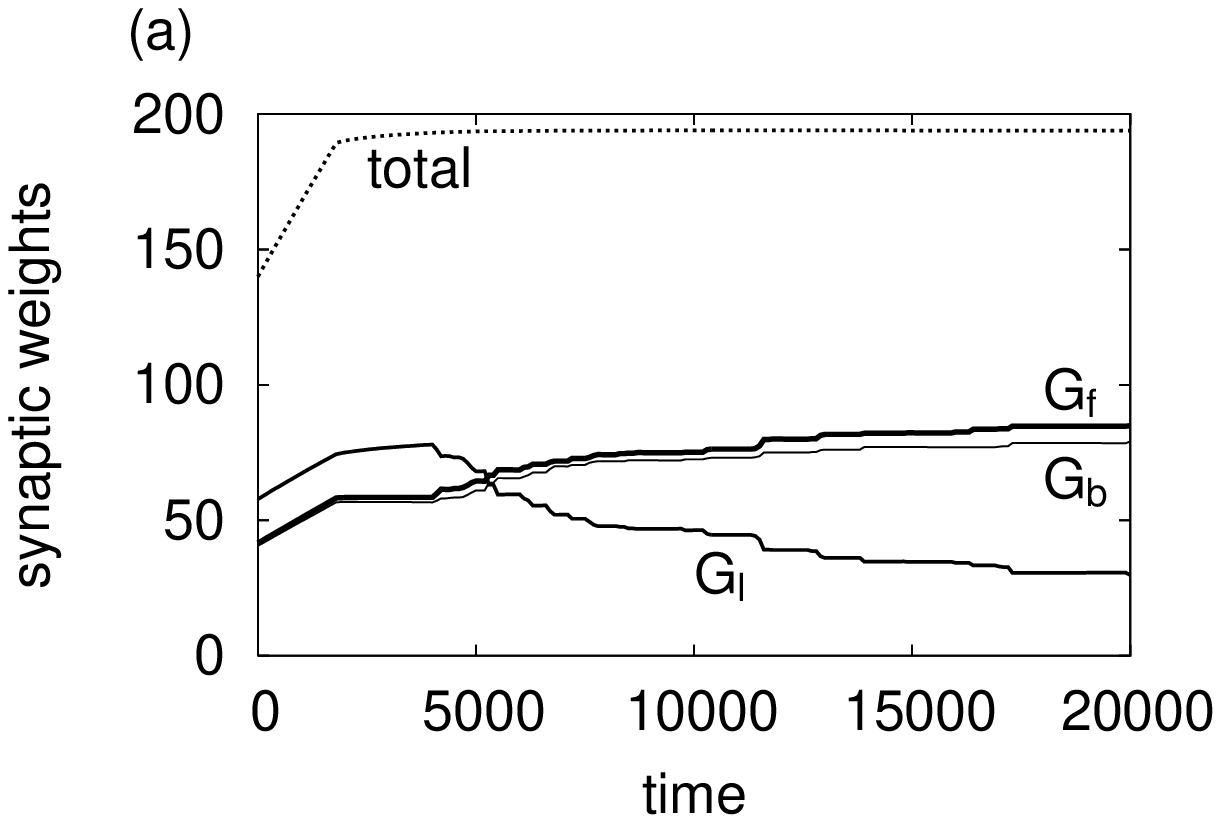}
\includegraphics[height=6.0cm,width=6.0cm]{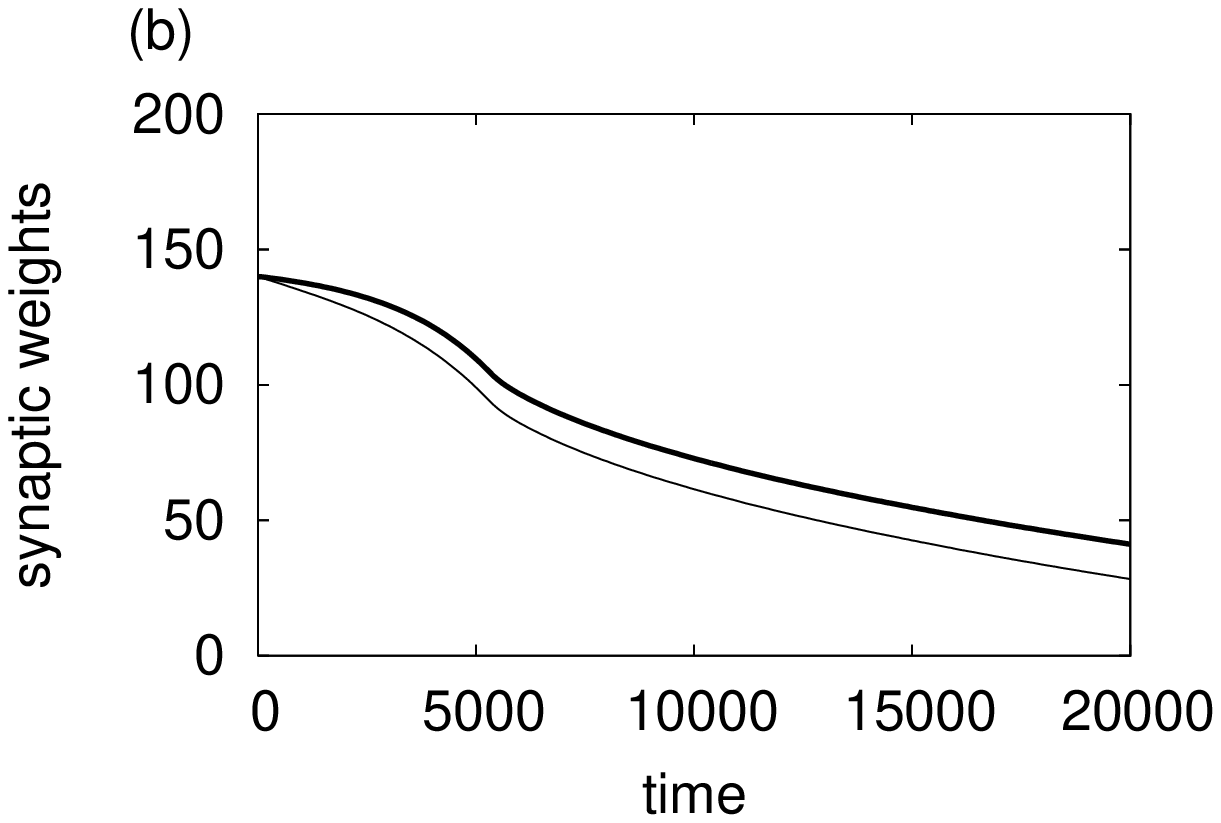}
\includegraphics[height=6.0cm,width=6.0cm]{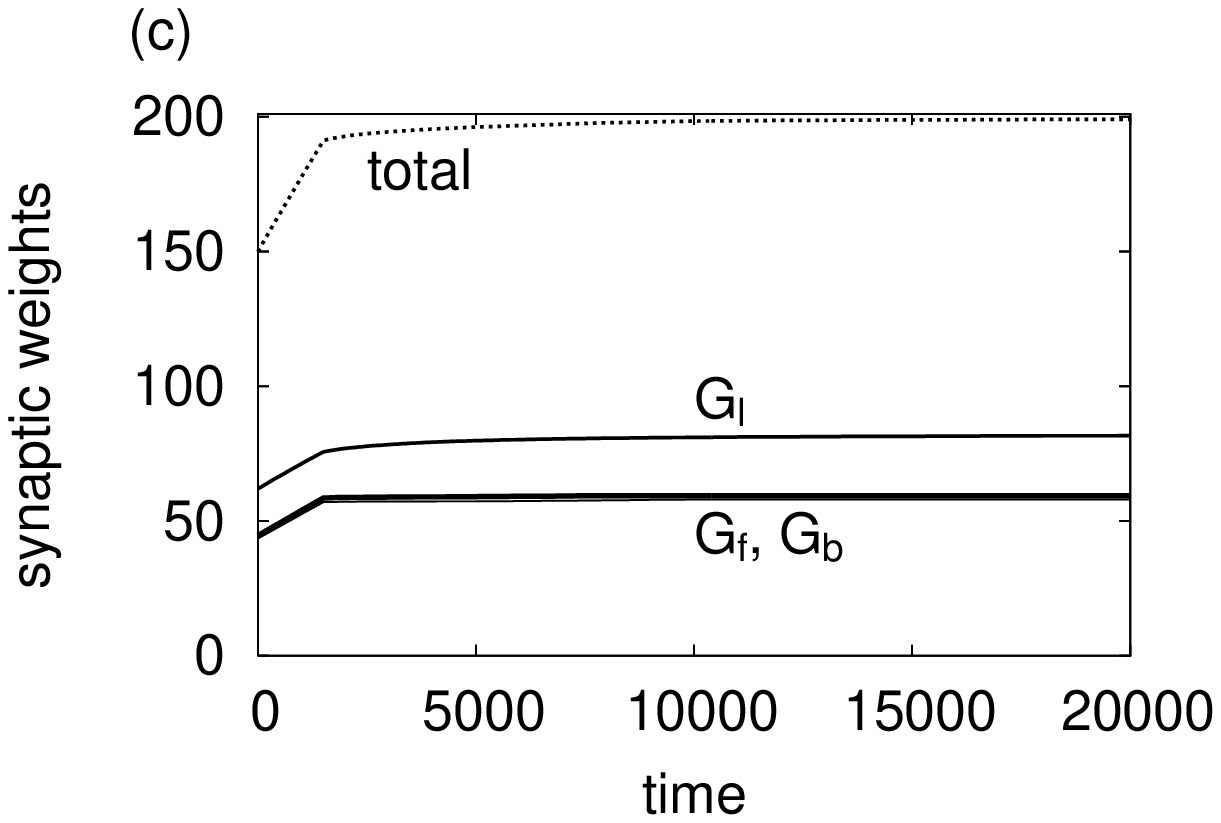}
\includegraphics[height=6.0cm,width=6.0cm]{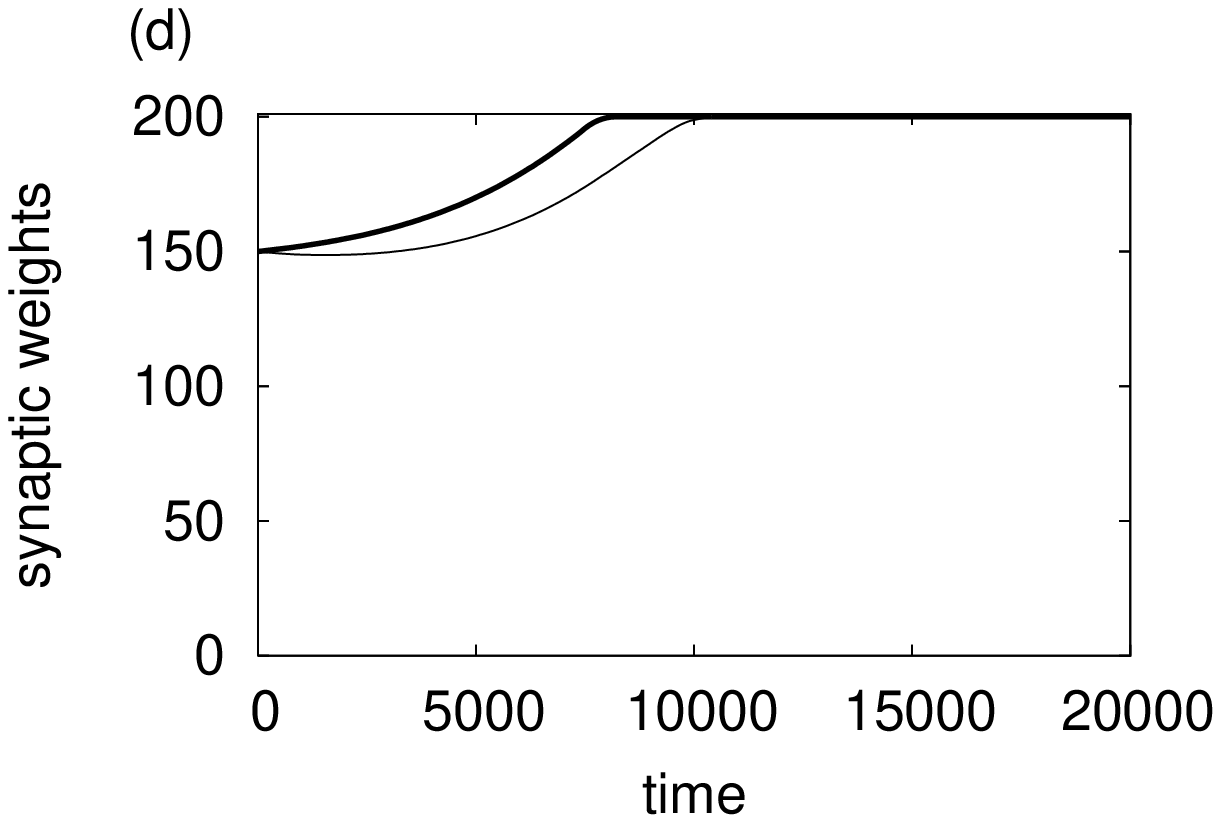}
\caption{Results for 100 randomly coupled oscillators subject to
symmetric STDP.  Evolution of the synaptic weights are shown for
$g(0)=140$ (a, b) and $g(0)=150$ (c, d). See the caption of
\FIG\ref{fig:asym} for legends. In (c), $G_f$ (thick solid line) and
$G_b$ (thin solid line) overlap almost completely.
Time courses of (e) $L$ and (f) $r$
are compared for $g(0)=140$, 145, 146, 148, and 150. 
In (e), lower lines correspond to larger $g(0)$.
(g) Final network structure for $g(0)=150$.}
\label{fig:sym}
\end{center}
\end{figure}

Now we examine symmetric STDP.
We set $g_{max}=200>g_c \cong 100.7$
so that frequency synchrony with small phase lags
is achieved if 
$g_{ji}=g_{max}$ for all $(j,i)\in E$.
For the network same as that used in \SEC\ref{sub:asym}, evolution of
synaptic weights are summarized in \FIGS\ref{fig:sym}(a, b) and (c, d) for
$g(0)=140$ and $g(0)=150$, respectively. Because the oscillators
share an identical natural frequency, the phases are fairly close
among them even with weak coupling.
The synapses among these oscillators are potentiated.
 Accordingly, for both values of
$g(0)$, the average synaptic 
weight initially increases
(dotted lines in \FIG\ref{fig:sym}(a, c)).  Note that the forward
weights ($G_f$, thick solid lines in \FIG\ref{fig:sym}(a, c))
and the backward weights ($G_b$,
thin solid lines in \FIG\ref{fig:sym}(a, c)) are equally potentiated.
Accordingly, 
the distance between the oscillators is initially shortened to result in
a decrease in $L$ (\FIG\ref{fig:sym}(e), $t \le 2000$).

When $g(0)=140$, the average synaptic weight stops increasing at a
value slightly smaller than $g_{max}=200$ (dotted line in
\FIG\ref{fig:sym}(a)).  This is because the synapses linking the
pacemaker to the oscillators have not been potentiated. Actually,
$G_f^0$ (thick line in \FIG\ref{fig:sym}(b)) and $G_b^0$ (thin line in
\FIG\ref{fig:sym}(b)) decrease to eventually decouple the oscillators
from the pacemaker. Consequently, $L$ diverges (uppermost
line in \FIG\ref{fig:sym}(e)), and frequency synchrony is eventually
lost (\FIG\ref{fig:sym}(f)).

When $g(0)=150$, $G_f^0$ (thick line in \FIG\ref{fig:sym}(d))
and $G_b^0$ (thin line in \FIG\ref{fig:sym}(d)) as well as 
$G_f$ (thick solid line in \FIG\ref{fig:sym}(c))
and $G_b$ (thin solid line in \FIG\ref{fig:sym}(c)) increase.
Consequently, $L$ continues to decrease to reach the minimum possible
value for which $g_{ji} = g_{max}$ is achieved for most synapses 
(lowermost line in \FIG\ref{fig:sym}(e)).  The synchrony measure $r$
stays near unity throughout (\FIG\ref{fig:sym}(f)).  In fact,
approximate 
phase synchrony as well as frequency synchrony has been realized quickly.
The synchrony arises not owing to STDP but
to sufficiently strong initial coupling.  

Based
on \FIGS\ref{fig:sym}(e) and (f), which show time courses of 
$L$ and $r$ for several values of $g(0)$,
we estimate $145< g_{c-stdp}< 146$.
This value of $g_{c-stdp}$ 
is much larger than the case of asymmetric STDP and 
comparable to $g_c \cong 100.7$, that is, the threshold
for frozen synapses.

The fact that symmetric STDP does not promote frequency synchrony
manifests the importance of the feedforwardness of networks.  In
general, forward synapses promote frequency synchrony, whereas
backward synapses hamper it \cite{Kori04}.  Symmetric STDP does not
independently control forward synaptic weights and backward synaptic
weights. Consequently, it cannot get rid of backward synapses without
sacrificing forward synapses.  Final network structure is shown for
$g(0)=150$ in \FIG\ref{fig:sym}(g). In contrast to the case of
asymmetric STDP (\FIG\ref{fig:asym}(g)), many backward synapses (thick
lines) remain. Under symmetric STDP, frequency
synchrony is due to strong mutual
interaction
but not to formation of a feedforward network.

\section{Networks of Pulse-coupled Spiking Neurons}\label{sec:iz}

To inspect whether the results obtained for coupled phase oscillators
qualitatively apply to
more general neuron models, we numerically simulate
pulse-coupled spiking neurons under STDP. We adopt a two-dimensional
neuron model \cite{Izhikevich03IEEE,Izhikevich04CC}. The subthreshold 
dynamics of the $i$-th neuron are described by
\begin{eqnarray}
\dot{v}_i &=& 0.04 v_i^2 + 5 v_i + 140 - u_i - I_{syn,i} - I_{ext,i},
\label{eq:iz_v}\\
\dot{u}_i &=& a (bv_i-u_i),
\label{eq:iz_u}
\end{eqnarray}
where $v_i$ is the membrane potential (mV), $u_i$ denotes the recovery
variable that evolves slowly relative to $v_i$, and 
the time unit is millisecond. The
spiking mechanism is implemented by
resetting the dynamical variables to $(v_i, u_i)=(c,d)$
as soon as $v_i$ exceeds 30 mV. We set $a=0.02$,
$b=0.2$, $c=-65$, and $d=8$, which are standard parameter values for
modeling pyramidal neurons \cite{Izhikevich03IEEE,Izhikevich04CC}.

\begin{figure}
\begin{center}
\includegraphics[height=6.0cm,width=6.0cm]{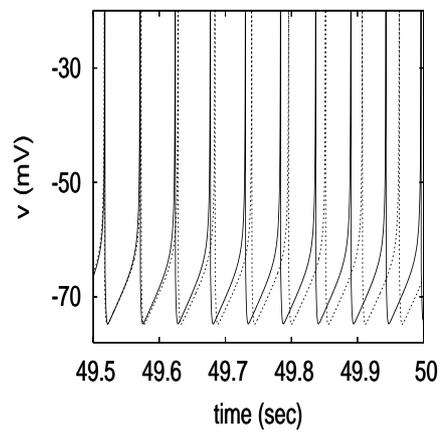}
\caption{Sample traces of $v_0$ (solid line) and $v_1$ (dashed line)
when the spiking neurons are uncoupled.}
\label{fig:iz_memb}
\end{center}
\end{figure}

The input $I_{ext,i}$ and $I_{syn,i}$ are the external bias input and
the synaptic input, respectively.  We set $I_{ext,0} = 8.4$ and
$I_{ext,i} = 8$ ($1\le i\le n-1$).
The inherent
firing rate of the pacemaker ($=18.8$ Hz) is about 5 \% higher than
that of the oscillators ($=17.9$ Hz). In \FIG\ref{fig:iz_memb},
example traces of the membrane potentials of the pacemaker (solid line)
and that of an oscillator (dashed line) are shown.

The synaptic input $I_{syn,i}$ is composed
of superposition of incident spikes from the neurons presynaptic
to the $i$-th neuron.  A presynaptic spike of the $j$-th neuron
($(j,i)\in E$) is assumed to change
the postsynaptic membrane potential according to the time course
given by the alpha function:
\begin{equation}
g_{ji} s(t) = g_{ji} \alpha^2 t e^{-\alpha t},\quad t\ge 0,
\end{equation}
where $t=0$ corresponds to the spike time.
We set $\alpha = 1$, so that
the unit 
synaptic input $s(t)$ peaks at $t=1/\alpha = 1$ ms 
and then decays slowly.
We set $A^+=0.09$, $A^- = 0.1$, and $\tau=10$ ms.

The random network with $n=100$ used in the following simulations are
the same as that used in \SEC\ref{sec:large}.
We set the initial synaptic weight $g_{ji} = g(0)$ for all the synapses.
The initial values of $v_i$ and $u_i$ are independently chosen according
to the uniform distributions on $[-75,-50]$ and $[-8,-6]$, respectively.
Under these conditions, we track time courses of $G_f$, $G_b$,
$G_l$, $G_f^0$, $G_b^0$, $L$ defined in
\SEC\ref{sub:order_parameters}, and $r=r([t_1,t_2])$ 
redefined based on spike counts:
\begin{equation}
r = \frac{\frac{1}{n-1}\sum^{n-1}_{i=1} \mbox{(number of spikes
from the $i$-th neuron)}}
{\mbox{(number of spikes from the pacemaker)}}.
\end{equation}
Frequency synchrony yields $r\cong 1$. If the pacemaker and the oscillators
fire independently, $r$ is the ratio of the single-neuron firing rate
to the pacemaker firing rate. We set $t_2-t_1= 10000$ ms. 

With these parameter values, we first determined $g_c$ without
STDP.
We numerically obtained $g_c \cong 1.6$
for the network of one pacemaker and one oscillator, and
$g_c\cong 38$ for the random
network. In 
the following simulations with
asymmetric STDP, we set $g_{max}=35 <
g_c$ so
that uniform increases in $g_{ji}$
do not cause synchronization. Frequency synchrony 
requires feedforward network structure.

\begin{figure}
\begin{center}
\includegraphics[height=6.0cm,width=6.0cm]{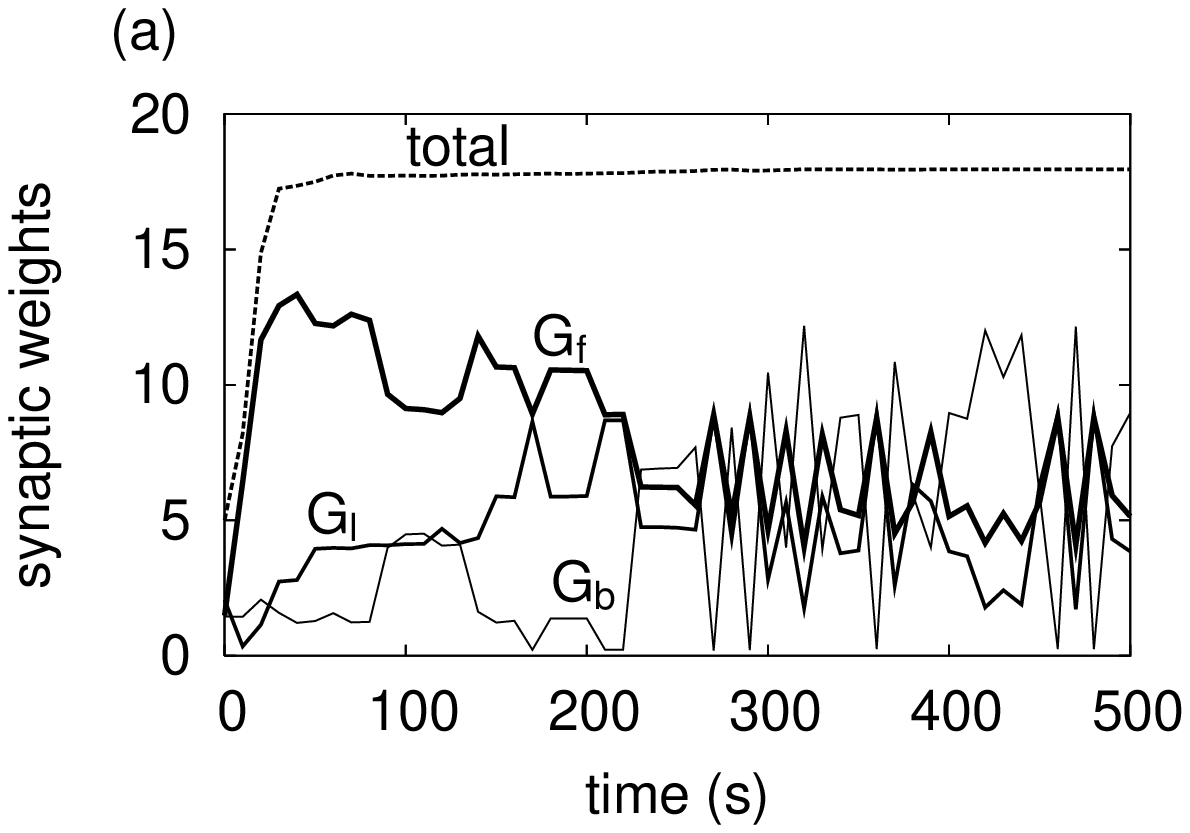}
\includegraphics[height=6.0cm,width=6.0cm]{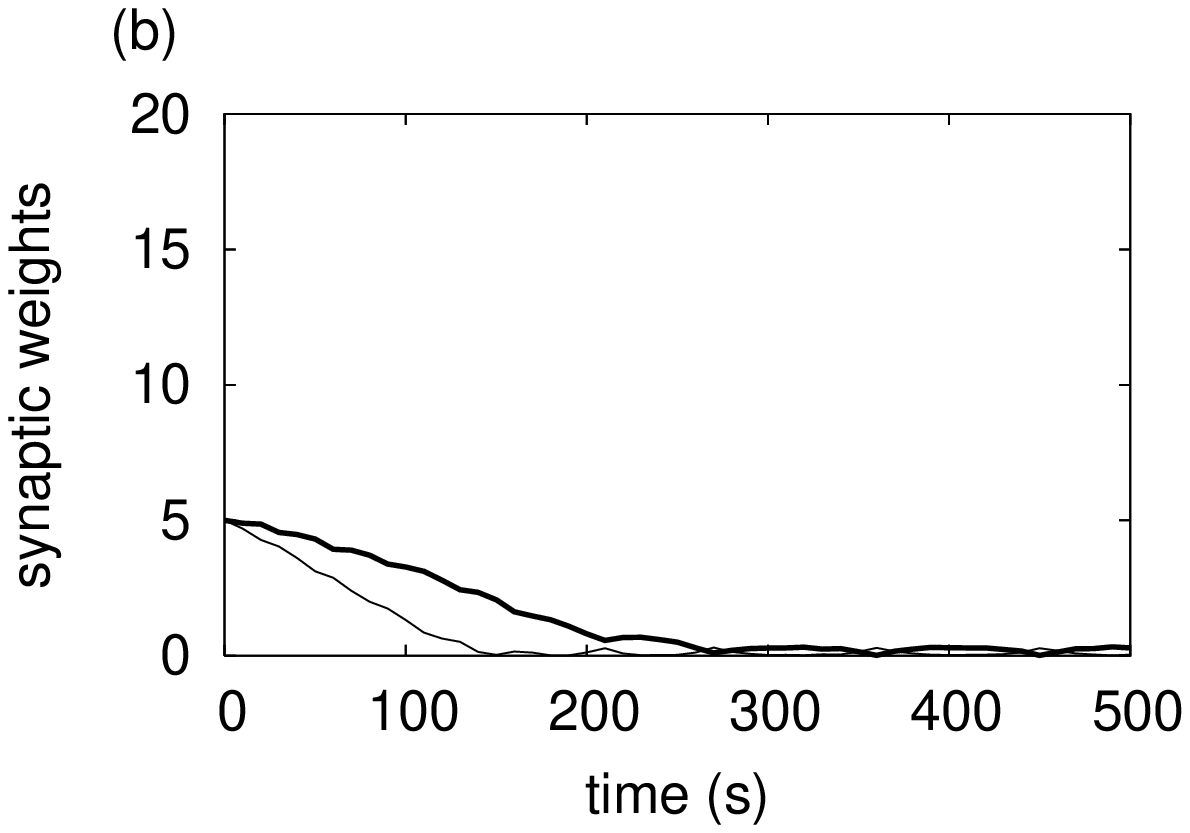}
\includegraphics[height=6.0cm,width=6.0cm]{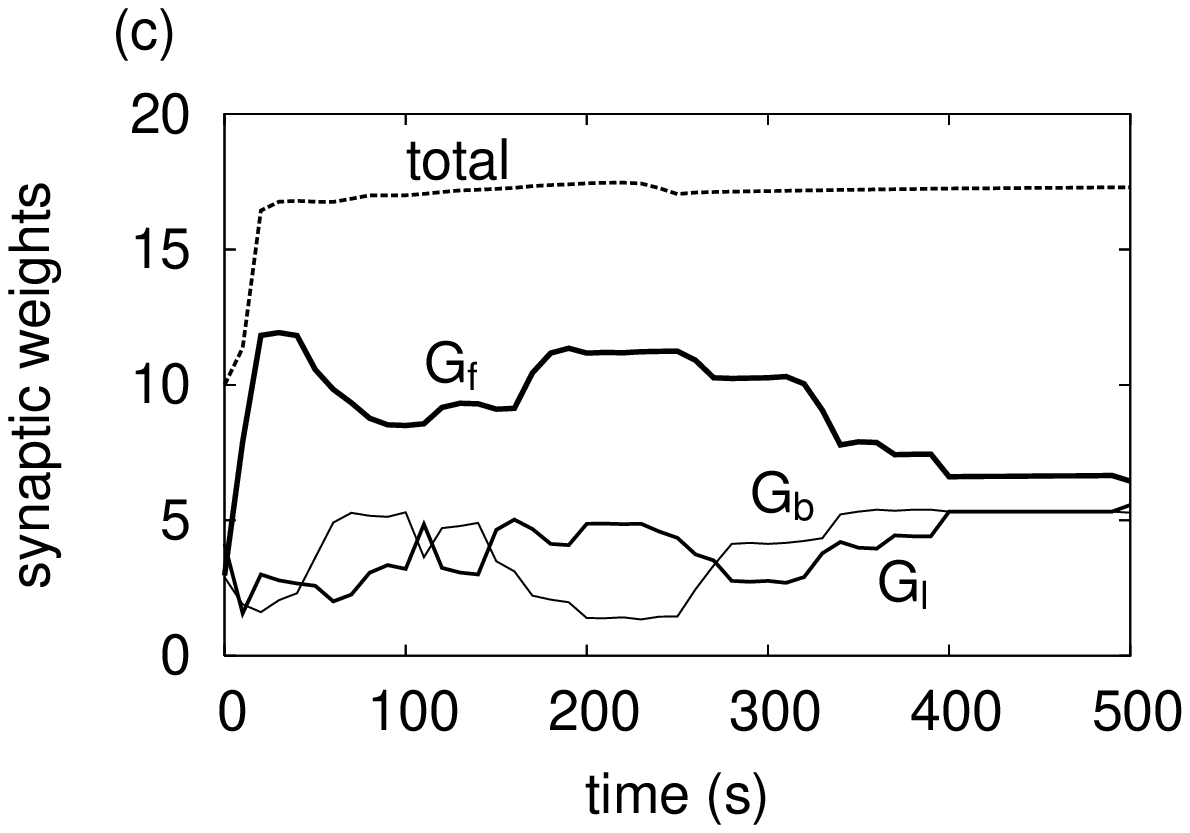}
\includegraphics[height=6.0cm,width=6.0cm]{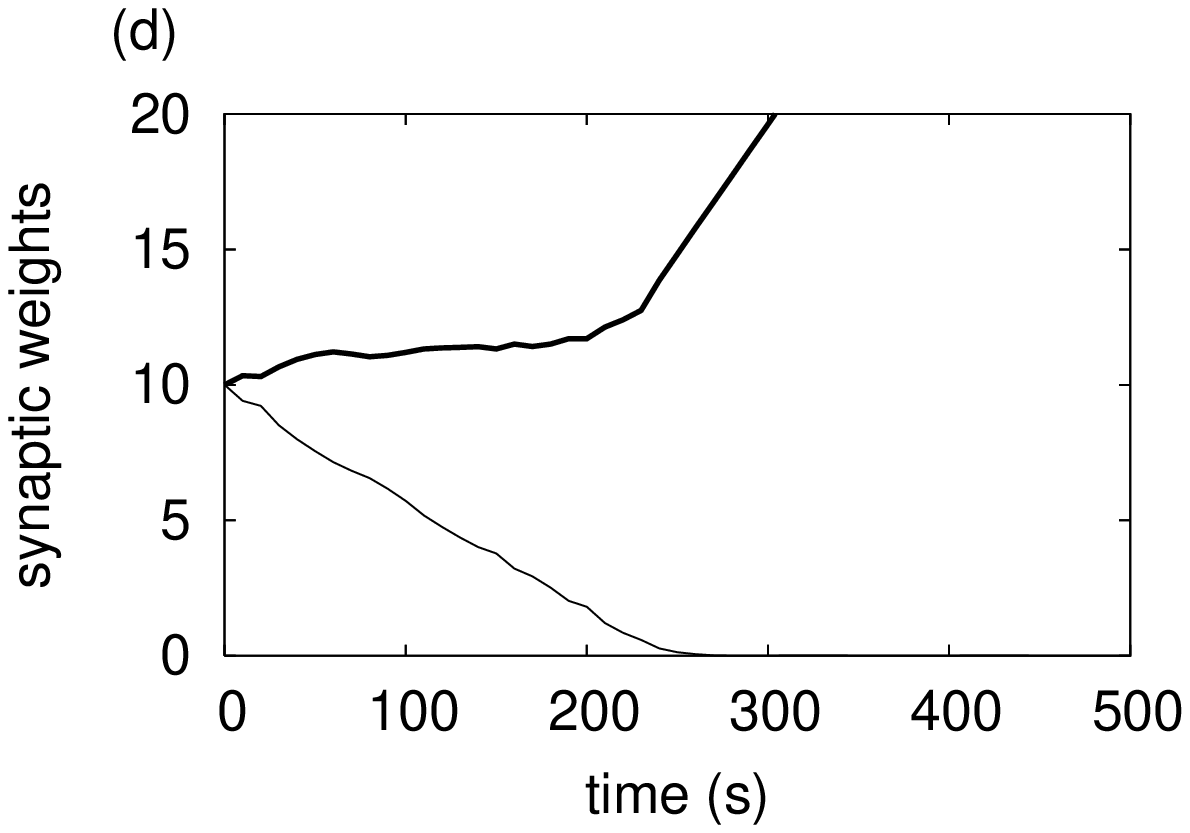}
\includegraphics[height=6.0cm,width=6.0cm]{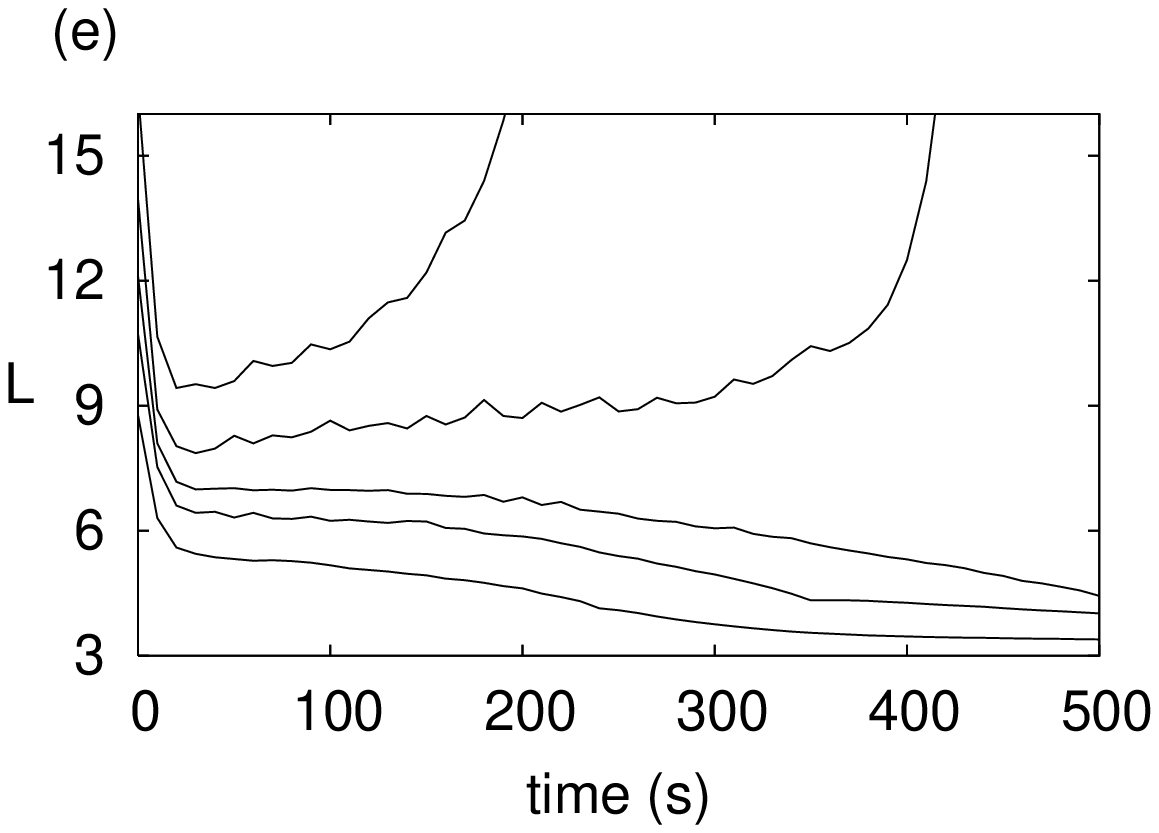}
\includegraphics[height=6.0cm,width=6.0cm]{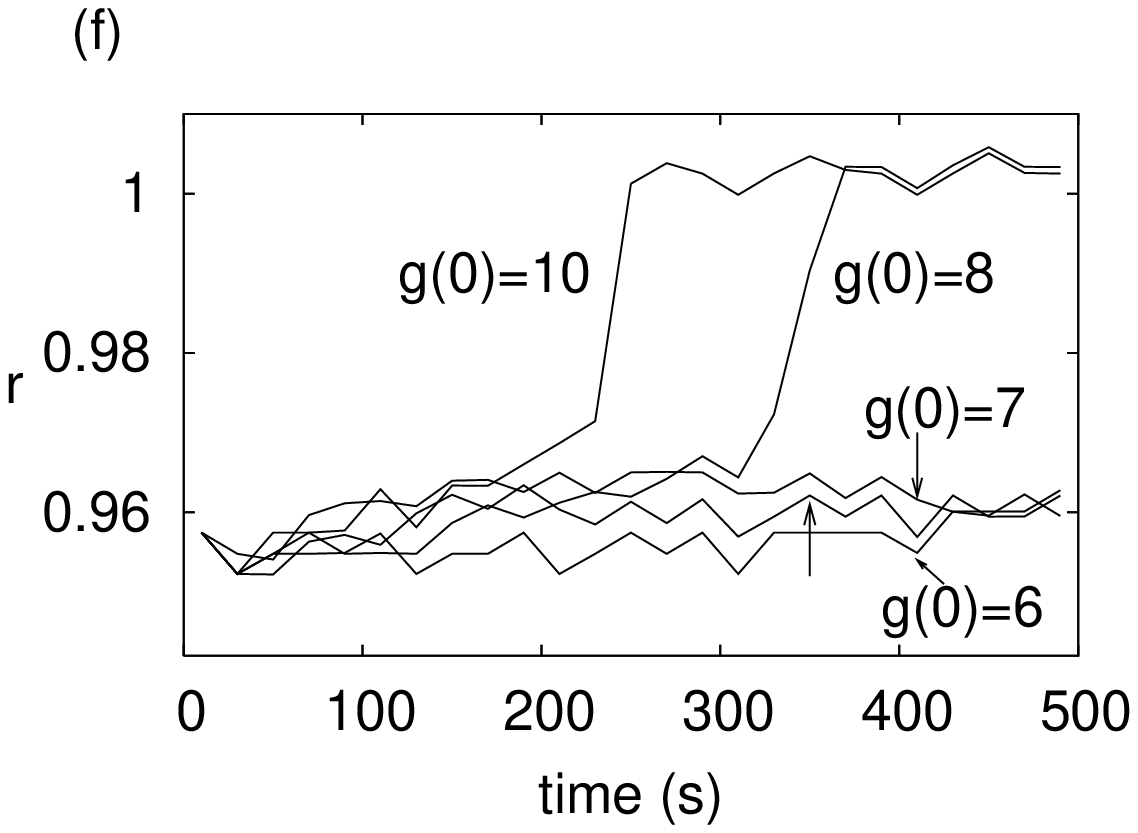}
\caption{Results for 100 randomly coupled
spiking neurons subject to asymmetric STDP.
Evolution of the synaptic weights 
are shown for (a, b) $g(0)=5$ and (c, d) $g(0)=10$.
See the caption of \FIG\ref{fig:asym} for legends. 
Time courses of (e) $L$ and (f) $r$
are compared for $g(0)=5$, 6, 7, 8, and 10.
In (e), lower lines correspond to larger $g(0)$.}
\label{fig:iz}
\end{center}
\end{figure}

For homogeneous initial synaptic
weights $g(0)=5$ and $g(0)=10$,
evolution of synaptic weights via asymmetric STDP 
is shown in
\FIGS\ref{fig:iz}(a, b) and \ref{fig:iz}(c, d), respectively. 
For $g(0)=5$, $G_f$ (thick solid line in \FIG\ref{fig:iz}(a)) and
$G_f^0$ (thick line in \FIG\ref{fig:iz}(b)) do not grow during the
course of plasticity, similar to \FIGS\ref{fig:asym}(a, b).
The oscillators disconnect from the pacemaker, and frequency synchrony
is not realized (\FIG\ref{fig:iz}(f)).  For
$g(0)=10$, the forward connection from the pacemaker to the set of
oscillators is established (thick line in \FIG\ref{fig:iz}(d)), and
backward connection is gradually removed (thin line in
\FIG\ref{fig:iz}(d)), similar to \FIGS\ref{fig:asym}(c, d). As a result,
frequency synchrony is reached (\FIG\ref{fig:iz}(f)).
Based on \FIGS\ref{fig:iz}(e) and
\ref{fig:iz}(f), which respectively show $L$ and $r$ for different
$g(0)$, we estimate $g_{c-stdp} \cong 7$, which is much smaller than 
$g_c\cong 38$.
Note that 
$g(0)=7$ is a marginal case, which yields a long transient
before frequency synchrony is reached. 
Frequency synchrony is not induced with, for example,
$g(0)=10<g_c$ if synapses are
frozen. These numerical
simulations confirm that the results derived in the
previous sections apply to networks of pulse-coupled spiking
neurons.

\section{Discussion}

We have shown that asymmetric STDP greatly reduces the threshold
for frequency synchronization of neural networks with a pacemaker.
This reduction is efficient particularly when LTP and
LTD are nearly balanced, as assumed for stabilization of
synaptic weights in previous literature
\cite{Kempter99,Song00,Song01,Vanrossum00}.  Our analytical results
for two-oscillator networks provide theoretical understanding of
STDP-induced synchrony of two-body networks with real neurons
\cite{Nowotny03JNS} and with Hodgkin-Huxley neurons
\cite{Zhigulin03}. Our numerical results for large networks extend
earlier numerical simulations \cite{Zhigulin04} and provide 
mechanisms of synchrony.

More microscopically, we have shown that STDP guides formation of
feedforward networks originating from the pacemaker
(\FIG\ref{fig:asym}(g)). By eliminating backward connection, frequency
synchrony is promoted in terms of required synaptic weights. Networks
self-organize by asymmetric STDP so that upstream neurons entrain
downstream neurons. 
Even though engineered
learning algorithms can
promote formation of feedforward networks
\cite{Kori06}, asymmetric STDP naturally achieves this goal.
Facilitation of frequency synchrony does not occur
for symmetric STDP, which cannot suppress backward synapses without
sacrificing forward synapses. 

In recurrent networks, synaptic delay may destabilize
otherwise stable synchrony, leading to oscillatory or chaotic
population dynamics (e.g. Gerstner and van Hemmen, 1993; Gerstner, 2000;
Timme et al., 2002). 
However, our numerical simulations suggest that
this is not the case in our system, which
can be explained as follows. In the phase oscillator model, the
effect of delay can be replaced in a good approximation by the phase
shift in the coupling function \cite{Kori01,Kori03}.
Dynamics of the oscillator system under
consideration do not change qualitatively for a large class of the
coupling function \cite{Kori06}, which effectively
includes the case of synaptic delay. Although a synaptic
delay enlarges the
phase difference between connected neurons, the oscillator
dynamics in our model are thus robust against delay. Another possible
complication is that synaptic delay may change synaptic evolution
because it could interact with the learning window. However, since
the delay simply increases 
the phase difference between connected neurons,
the causality of spike timing does not change even with
delay, as corroborated by our numerical experiments.

In terms of network structure, the feedforward structure is distinct
from pruning of synapses in a predefined unidirectional network with
many presynaptic neurons projecting to a single postsynaptic neuron
\cite{Kempter99,Song00,Song01,Vanrossum00}. It also differs from
multipartite networks each part of which forms a cluster of
synchronously firing neurons \cite{Horn,Levy}. In a sense, feedforward
structure and hierarchy are straightforward consequences of asymmetric
STDP \cite{Song01}, which opts for causality. We stress that
feedforward structure is naturally organized when
a network has a distinct pacemaker. The idea of growing feedforward
structure by unsupervised learning dates back to pioneering work by
Bienenstock (1991, 1995), which employed Hebbian plasticity.
We have analyzed the network formation 
in detail under asymmetric STDP.
In real neural networks,
there may be multiple pacemakers as well as a huge number of follower
neurons.  It is straightforward to extend our results to the case in
which a collection of neurons in a network serves as a pacemaker.
Pacemaker neurons are relevant to, for example, regulation of
respiration, internal clock, and Parkisonian diseases (see
\SEC\ref{sec:introduction} for references). In these brain regions,
pacemaker neurons may recruit downstream neurons for frequency
synchrony in order to, for example,
amplify rhythmic activity.

Formation of feedforward
structure could occur even when predetermined 
pacemakers are absent.  In this
case, neurons with relatively high natural
frequencies may play a role of pacemaker.  Backward connection to these
fast neurons, which would perturb their periodic firing,
can be eventually eliminated by asymmetric STDP. Then,
the fast neurons can serve as distinct pacemakers.
Regardless of the initial
presence of pacemakers, asymmetric STDP creates
frequency synchrony, which can be called
{\it feedforward synchrony}.
This mechanism of synchrony differs from that of
synchrony based on mutual coupling \cite{Kuramotobook}.

Our results do not suggest that asymmetric STDP promotes phase
synchrony, namely, spike synchrony.  This is in contrast to the finding
that phase synchrony is caused by asymmetric STDP \cite{Karbowski}.
In their work, fixed inhibitory coupling as well as excitatory
coupling with asymmetric
STDP was assumed. Perfectly balanced LTP and LTD, at
least as the average, is a key condition for the maintenance of
bidirectional connectivity and phase synchrony.  By contrast, we
assumed that LTD is stronger than LTP.  This yields a
considerable decrease in the threshold for synchrony. However, this
synchrony is frequency synchrony but not
spike synchrony.

With symmetric STDP, 
neurons whose
spike times are close are likely to bind together. Then, in addition
to frequency synchrony, approximate 
phase synchrony whose time resolution is
specified by the width of the learning window can develop
\cite{Seliger}. In some situations,
neurons divide into clusters in each of which rough spike
synchrony is maintained \cite{MasudaSTDP}. Unlike asymmetric
STDP, symmetric STDP does not lessen the threshold for synchrony.

When feedforward frequency synchrony is achieved, neurons at different
distances from the pacemaker fire asynchronously.  On top of that,
phase synchrony can be observed for neurons receiving the common
signal.  For example, the neurons directly connected to the pacemaker
are excited by the common drive from the pacemaker, so that
they are synchronized in phase.  Likewise,
neurons with the same distance from the pacemaker tend to fire
simultaneously. Indeed,
\FIG\ref{fig:raster}(b) and its magnification in
\FIG\ref{fig:raster}(c) indicate that clusters of phase-synchronized
neurons can be aligned according to the distance from the pacemaker 
\cite{Kori04}.
Even though spike time difference between neurons with
different distances is usually small, the order of firing is fixed and
reproducible. The pacemaker triggers a volley of spikes, which
travels down the hierarchy delineated by the distance.
This phenomenon is consistent with propagation of
synfire volley through feedforward neural networks in the
excitable regime \cite{Bienenstock95,Diesmann,Reyes,Vogels}.  However,
stably embedding synfire volley in recurrent networks is usually
difficult \cite{Mehring}.  It needs, for example, selective
enhancement of forward synapses by 10-fold, which corresponds to large
evoked EPSPs of 8 mV \cite{Vogels}.  With asymmetric STDP, forward
synapses are enhanced.  In addition, automatic elimination of backward
synapses appreciably lessens the forward synaptic strength (or the
size of EPSP) needed for stable synfire volley.

We have shown that the facilitation of frequency synchrony by
STDP is robust against some heterogeneity in the inherent
firing frequency of the neurons.
If synaptic
delays or neurons are strongly heterogeneous, we would obtain more complex
but reproducible spatiotemporal spike patterns
\cite{Izhikevich04CC,Izhikevich06NC,Lengyel}.

Oscillatory neurons can model,
for example, temporal coding of place cells
\cite{Mehta02} and hippocampal associative memory \cite{Lengyel}.
By contrast, many neural circuits
operate in the excitable regime, in which neurons are not spontaneously
oscillatory. 
Investigation of the excitable case is
warranted for future studies. However,
we believe that the conclusion that asymmetric
STDP but not symmetric STDP induces feedforward synchrony generalizes
to the excitable case, as is consistent with
previous numerical work
\cite{Song01,Zhigulin04}.

\section*{Acknowledgments}

We thank Brent Doiron and Taro Toyoizumi
for critical reading of the
manuscript and Tohru Ikeguchi and Tomoya Suzuki for discussion.
Naoki Masuda thanks the Special Postdoctoral Researchers
Program of RIKEN. Hiroshi Kori thanks financial support from the
Humboldt foundation (Germany) and from 21st Century COE program 
``Nonlinearity via Singularity'' in
Department of Mathematics, Hokkaido University.

\newpage
\pagestyle{empty}

\clearpage

\clearpage

\clearpage

\clearpage


\clearpage

\begin{center}
\includegraphics[height=6.0cm,width=6.0cm]{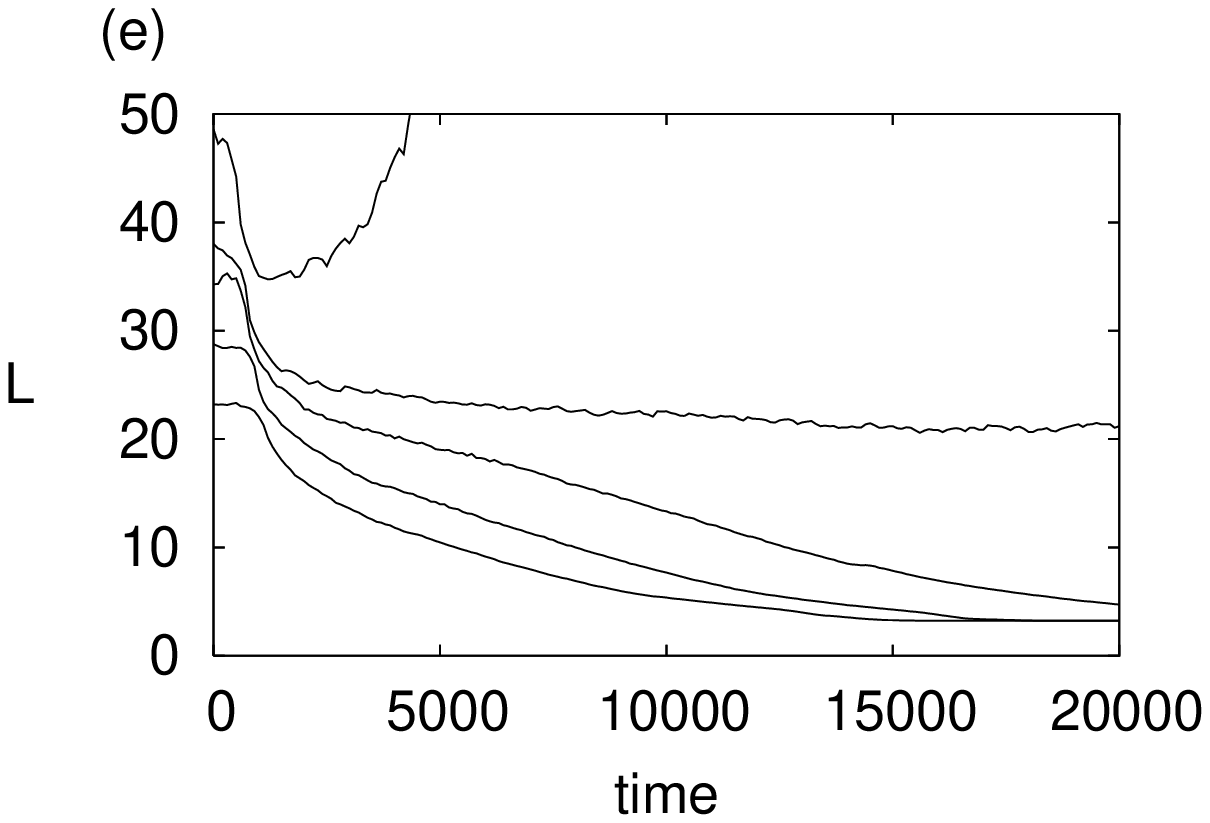}
\includegraphics[height=6.0cm,width=6.0cm]{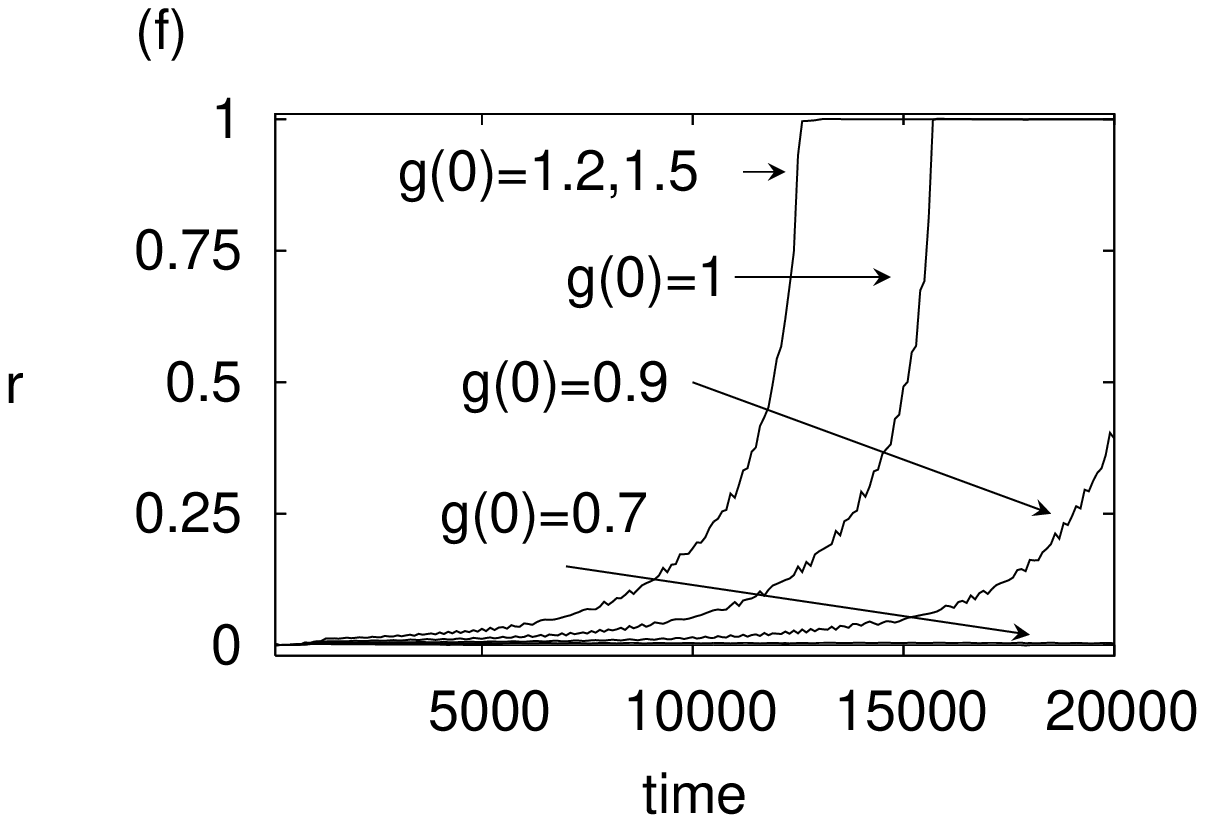}
\includegraphics[height=10.0cm,width=10.0cm]{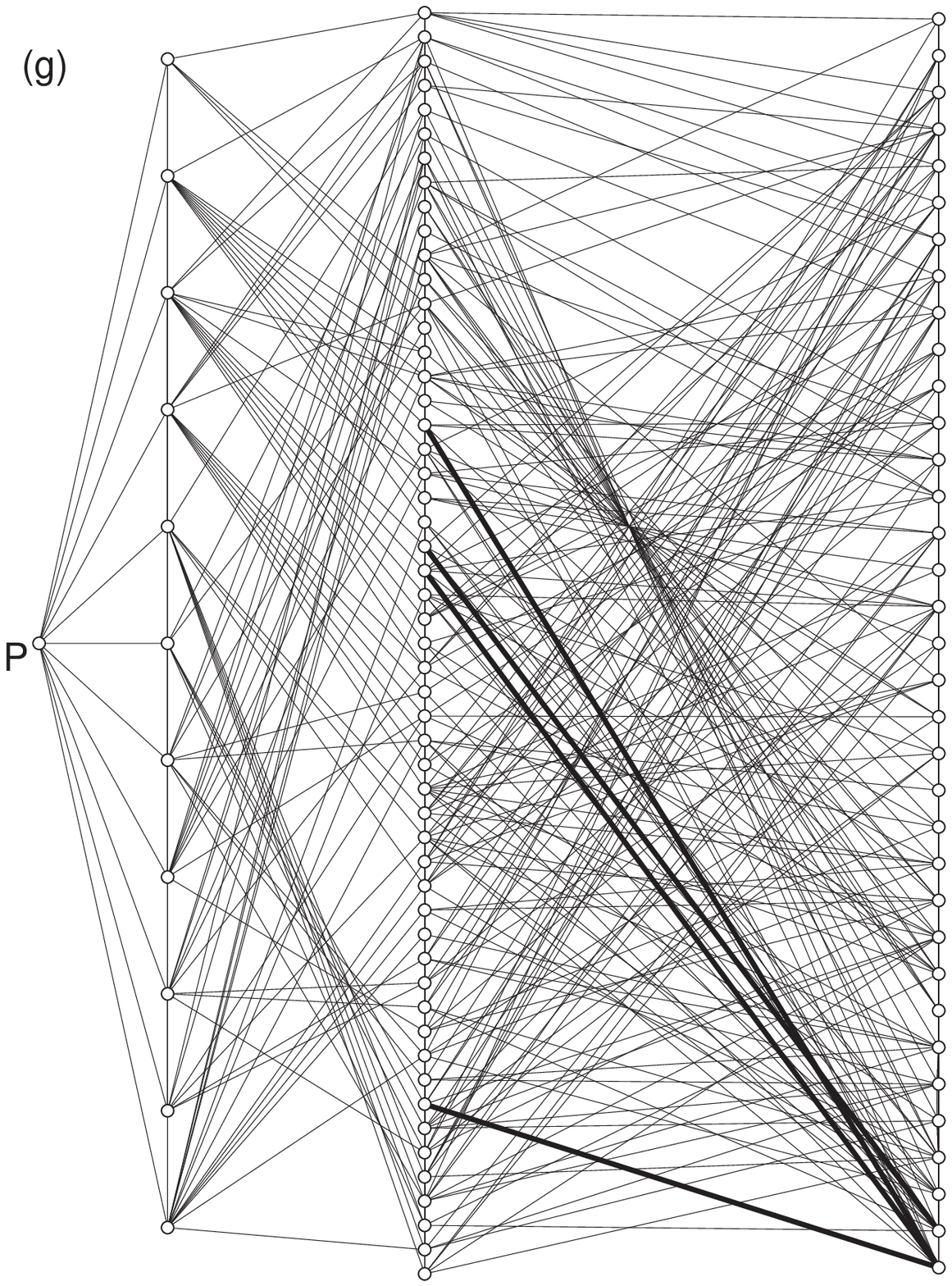}
\includegraphics[height=6.0cm,width=6.0cm]{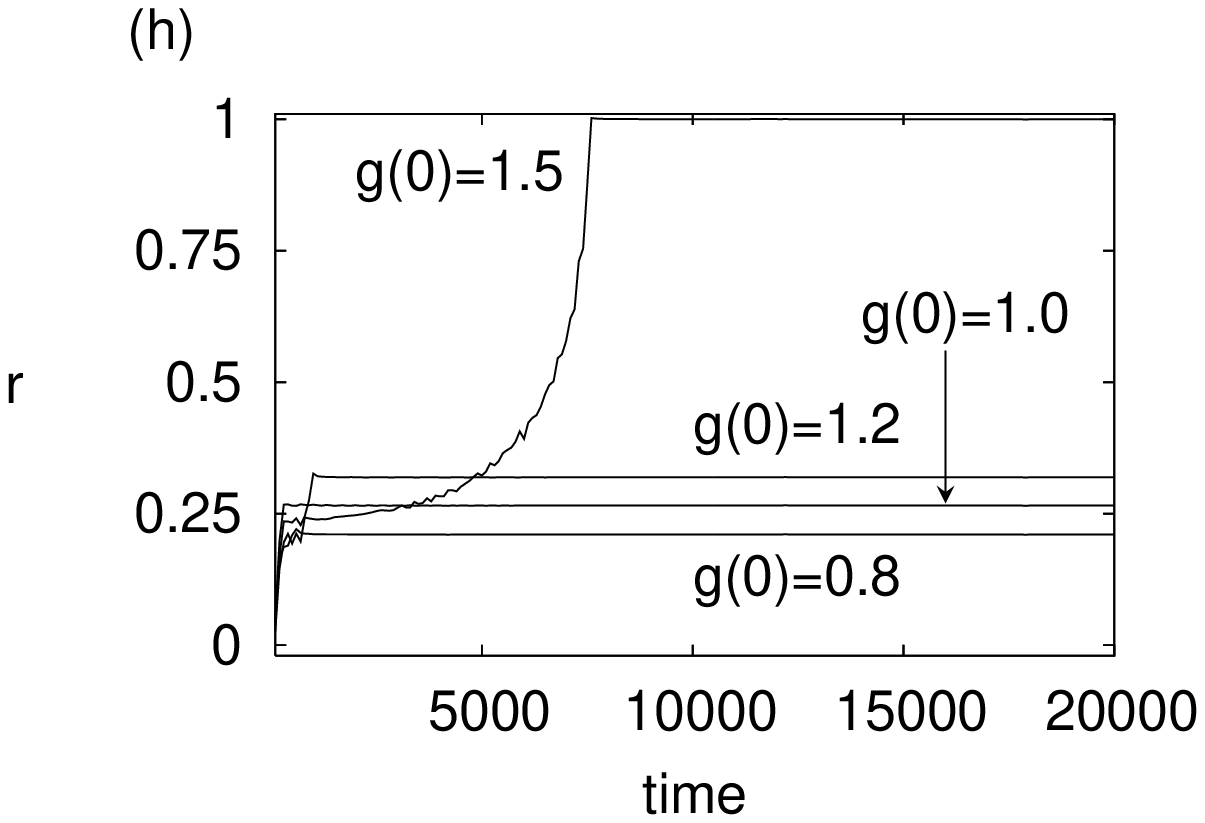}
\end{center}

\clearpage

\clearpage

\clearpage

\begin{center}
\includegraphics[height=6.0cm,width=6.0cm]{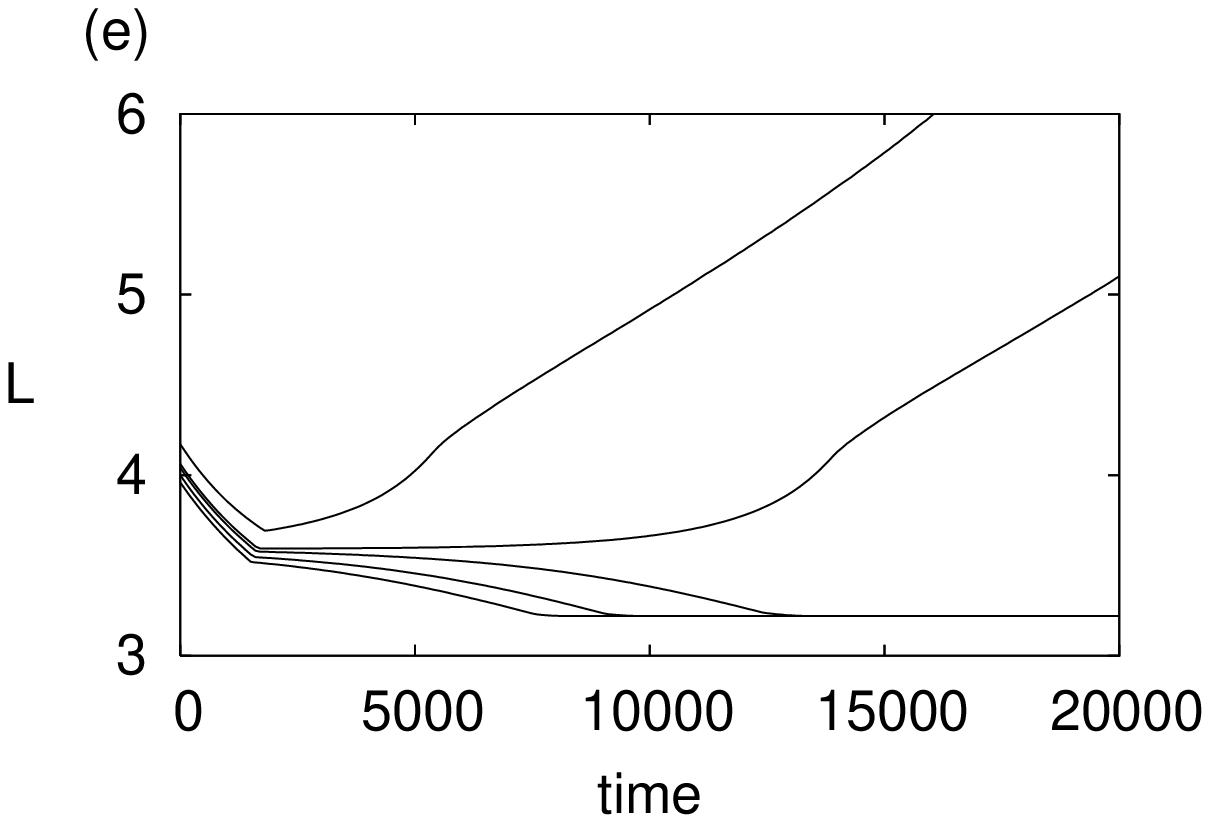}
\includegraphics[height=6.0cm,width=6.0cm]{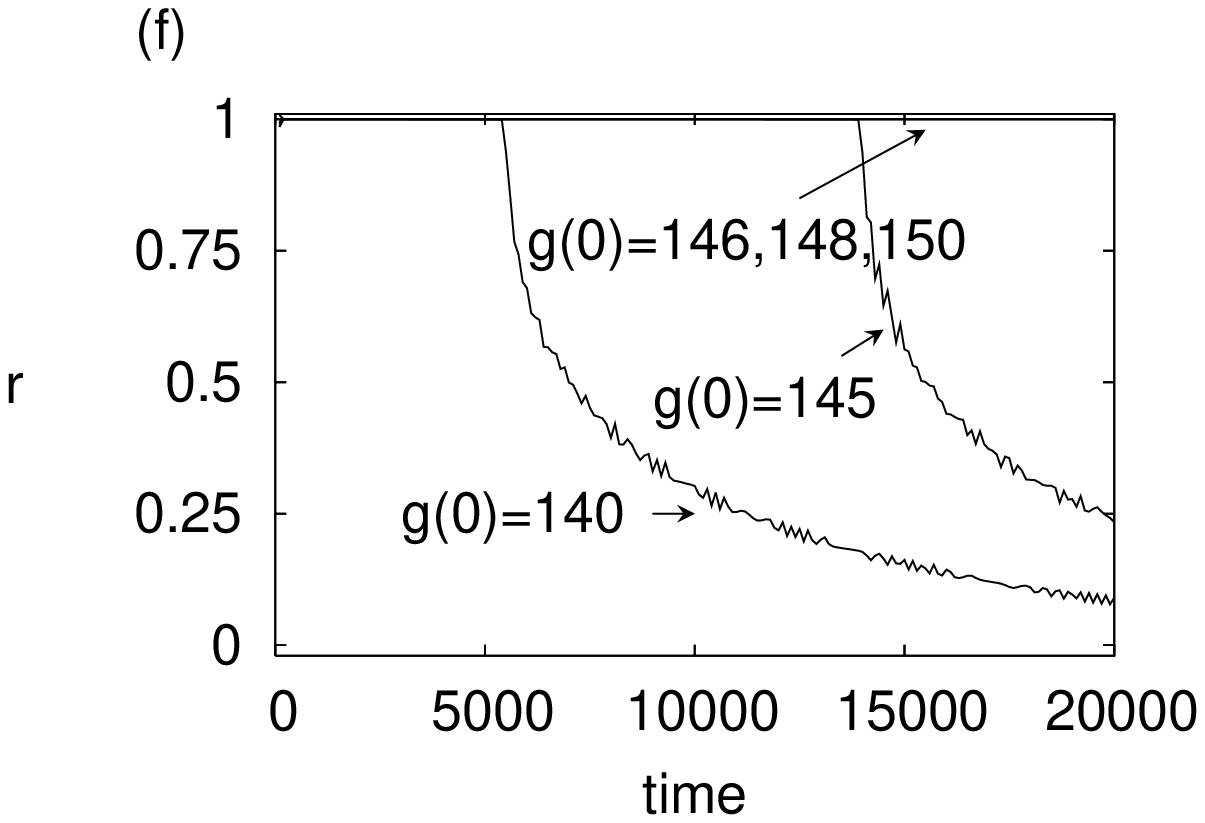}
\includegraphics[height=10.0cm,width=10.0cm]{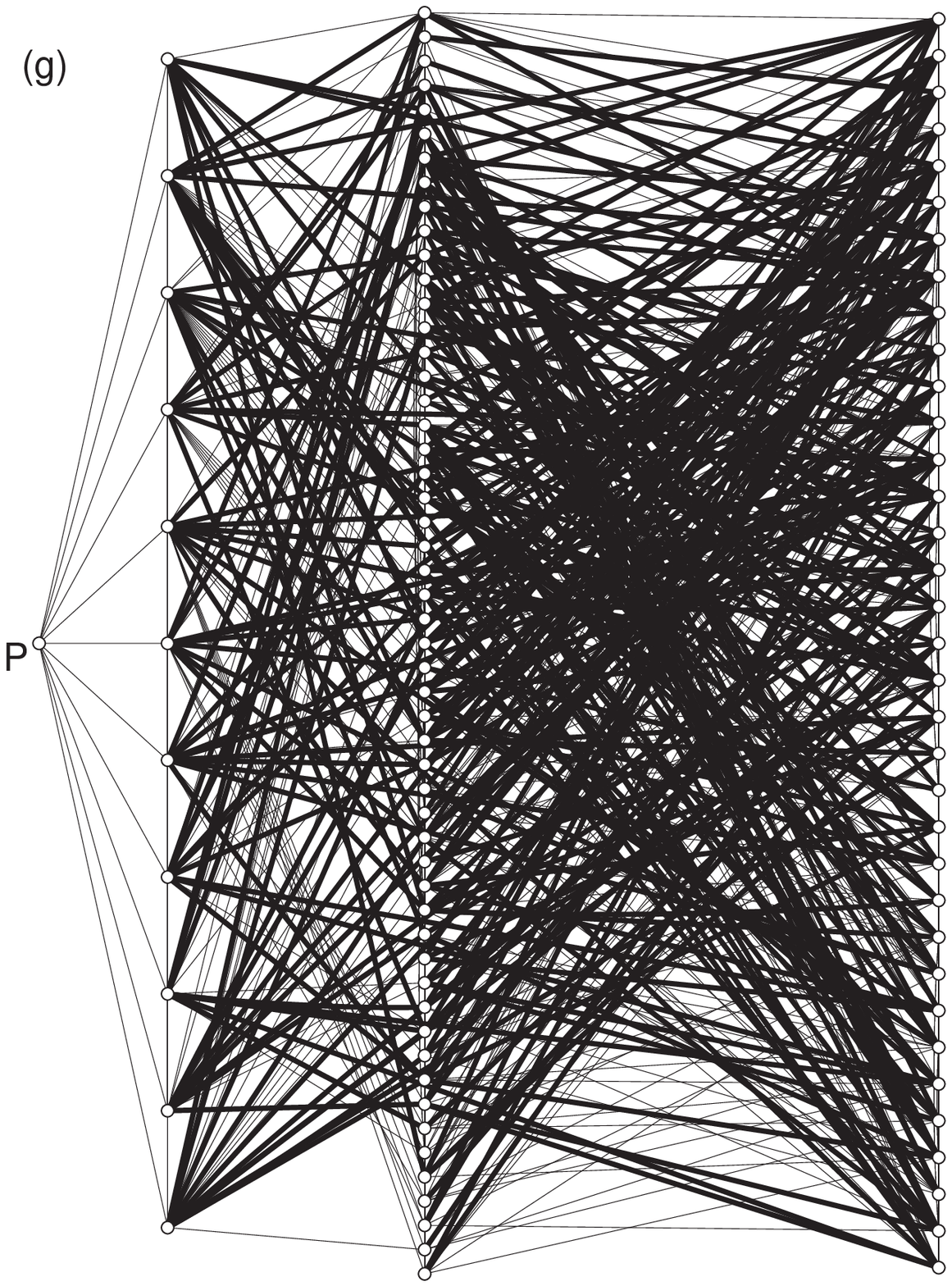}
\end{center}

\clearpage

\clearpage

\begin{thebibliography}{99}

\bibitem[Abbott and Nelson, 2000]{Abbott00}
Abbott LF, Nelson SB (2000)
Synaptic plasticity: taming the beast.
{\it Nat. Neurosci. Supp.} 3:1178--1183.

\bibitem[Bienenstock, 1991]{Bienenstock91}
Bienenstock E (1991) Notes on the growth of a composition machine.
Proceedings of the First Interdisciplinary Workshop on
Compositionality in Cognition and Neural Networks. Abbaye de
Royaumont, France (eds. Andler D,
Bienenstock E, Laks B): 25--43.

\bibitem[Bienenstock, 1995]{Bienenstock95}
Bienenstock E (1995) A model of neocortex. {\it Network: Computation
in Neural Systems} 6:179--224.

\bibitem[Braunstein et al., 2003]{Braunstein03}
Braunstein LA, Buldyrev SV, Cohen R, Havlin S, Stanley HE (2003)
Optimal paths in disordered complex networks.
{\it Phys. Rev. Lett.} 91:168701.

\bibitem[Bell et al., 1997]{Bell97}
Bell CC, Han VZ, Sugawara Y, Grant K (1997)
Synaptic plasticity in a cerebellum-like structure
depends on temporal order.
{\it Nature} 387:278--281.

\bibitem[Bi and Poo, 1998]{Bi98}
Bi G, Poo M (1998)
Synaptic modifications in cultured hippocampal neurons:
dependence on spike timing, synaptic strength, and postsynaptic cell type.
{\it J. Neurosci.} 18(24):10464--10472.

\bibitem[Buzs\'{a}ki and Draguhn, 2004]{Buzsaki04}
Buzs\'{a}ki G, Draguhn A (2004) Neuronal
oscillations in cortical networks.
{\it Science} 304:1926--1929.

\bibitem[Dan and Poo, 1992]{Dan}
Dan Y, Poo M (1992) Hebbian depression of isolated
neuromuscular synapses in vitro. {\it Science} 256:1570--1573.

\bibitem[Diesmann et al., 1999]{Diesmann}
Diesmann M, Gewaltig M-O, Aertsen A (1999)
Stable propagation of
synchronous spiking in cortical neural networks.
{\it Nature} 402:529--533.

\bibitem[Froemke and Dan, 2002]{Froemke02}
Froemke RC, Dan Y (2002)
Spike-timing-dependent synaptic
modification induced by natural spike trains.
{\it Nature} 416:433--438.

\bibitem[Gerstner and van Hemmen, 1993]{Gerstner93prl}
Gerstner W, van Hemmen JL (1993)
Coherence and incoherence in a globally coupled ensemble of pulse-emitting units. {\it Phys. Rev. Lett.} 71:312--315.

\bibitem[Gerstner et al., 1996]{Gerstner96nat}
Gerstner W, Kempter R, van Hemmen JL, Wagner H (1996)
A neuronal learning rule for sub-millisecond temporal coding.
{\it Nature} 383:76--78.

\bibitem[Gerstner, 2000]{Gerstner00}
Gerstner W (2000) Population dynamics of spiking neurons: fast
transients, asynchronous states, and locking. {\it Neural Comput.} 
12:43--89.

\bibitem[Gerstner and Kistler, 2002]{Gerstnerbook}
Gerstner W, Kistler WM (2002) Spiking neuron models.
Cambridge University Press, Cambridge.

\bibitem[Glass and Mackey, 1988]{Glassbook}
Glass L, Mackey MC (1988) From Clocks to Chaos -- the Rhythms
of Life. Princeton University Press, Princeton.

\bibitem[Hansel et al., 1993]{Hansel93}
Hansel D, Mato G, Meunier C (1993)
Phase dynamics for weakly coupled Hodgkin-Huxley neurons.
{\it Europhys. Lett.} 23(5):367--372.

\bibitem[Hansel et al., 1995]{Hansel95}
Hansel D, Mato G, Meunier C (1995)
Synchrony in excitatory neural networks.
{\it Neural Comput.} 7:307--337.

\bibitem[Horn et al., 2000]{Horn}
Horn D, Levy N, Meilijson I, Ruppin E (2000) 
Distributed synchrony of spiking neurons in a Hebbian cell assembly.
In S. A. Solla, T. K. Leen, \& K. -R. M\"{u}ller (Eds.),
{\it Advances in
Neural Information Processing Systems} 12:129--135.
MIT Press, Cambridge, MA.

\bibitem[Hutcheon and Yarom, 2000]{Hutcheon00}
Hutcheon B, Yarom Y (2000)
Resonance, oscillation and the intrinsic frequency preferences of
neurons.
{\it Trends in Neurosci.} 23(5):216--222.

\bibitem[Izhikevich, 2003]{Izhikevich03IEEE}
Izhikevich EM (2003)
Simple model of spiking neurons.
{\it IEEE Trans. Neur. Netw.} 14(6):1569--1572.

\bibitem[Izhikevich et al., 2004]{Izhikevich04CC}
Izhikevich EM, Gally JA, Edelman GM (2004)
Spike-timing dynamics of neuronal groups.
{\it Cereb. Cort.} 14(8):933--944.

\bibitem[Izhikevich, 2006]{Izhikevich06NC}
Izhikevich EM (2006)
Polychronization: computation with spikes.
{\it Neural Comput.} 18:245--282.

\bibitem[Jefferys et al., 1996]{Jefferys}
Jefferys JGR, Traub RD, Whittington MA (1996)
Neuronal networks for induced `40 Hz' rhythms.
{\it Trends in Neurosci.} 19(5):202--208.

\bibitem[Karbowski and Ermentrout, 2002]{Karbowski}
Karbowski J, Ermentrout GB (2002)
Synchrony arising from a balanced synaptic plasticity in a network
of heterogeneous neural oscillators.
{\it Phys. Rev. E} 65:031902.

\bibitem[Kempter et al., 1999]{Kempter99}
Kempter R, Gerstner W, van Hemmen JL (1999) 
Hebbian learning and
spiking neurons. {\it Phys. Rev. E} 59:4498--4514.

\bibitem[Kori and Kuramoto, 2001]{Kori01}
Kori H, Kuramoto Y (2001).
Slow switching in globally coupled oscillators: 
robustness and occurrence through delayed coupling.
{\it Phys. Rev. E} 63:046214.

\bibitem[Kori, 2003]{Kori03}
Kori H (2003)
Slow switching in a population of delayed pulse-coupled oscillators.
{\it Phys. Rev. E} 68:021919.

\bibitem[Kori and Mikhailov, 2004]{Kori04}
Kori H, Mikhailov AS (2004)
Entrainment of randomly coupled oscillator networks by a pacemaker.
{\it Phys. Rev. Lett.} 93:254101.

\bibitem[Kori and Mikhailov, 2006]{Kori06}
Kori H, Mikhailov AS (2006)
Strong effects of network architecture in the entrainment 
of coupled oscillator systems. {\it Phys. Rev. E} 74:066115.

\bibitem[Kuramoto, 1984]{Kuramotobook}
Kuramoto Y (1984). Chemical oscillations, waves,
and turbulence. Springer-Verlag, Berlin.

\bibitem[Kuramoto, 1991]{Kuramoto91}
Kuramoto Y (1991)
Collective synchronization of pulse-coupled
oscillators and excitable units.
{\it Physica D} 50:15--30.

\bibitem[Lengyel et al., 2005]{Lengyel}
Lengyel M, Kwag J, Paulsen O, Dayan P (2005)
Matching storage and recall: hippocampal spike timing-dependent
plasticity and phase response curves.
{\it Nat. Neurosci.} 8:1677--1683.

\bibitem[Levy et al., 2001]{Levy}
Levy N, Horn D, Meilijson I, Ruppin E (2001) Distributed
synchrony in a cell assembly of spiking neurons. {\it Neural Networks}
14:815--824.

\bibitem[Markram et al., 1997]{Markram97}
Markram H, L\"{u}bke J, Frotscher M, Sakmann B
(1997) Regulation of synaptic efficacy by coincidence of
postsynaptic APs and EPSPs. {\it Science} 275:213--215.

\bibitem[Masuda and Aihara, 2004]{MasudaSTDP}
Masuda N, Aihara K (2004).
Self-organizing dual coding based on spike-time-dependent plasticity.
{\it Neural Comput.} 16:627--663.

\bibitem[Mehring et al., 2003]{Mehring}
Mehring C, Hehl U, Kubo M, Diesmann M, Aertsen A (2003)
Activity dynamics and propagation of synchronous spiking in locally
connected random networks.
{\it Biol. Cybern.} 88:395--408.

\bibitem[Mehta et al., 2002]{Mehta02}
Mehta MR, Lee AK, Wilson MA (2002)
Role of experience and oscillations in transforming a rate code into a
temporal code. {\it Nature} 417:741--746. 

\bibitem[Nishiyama et al., 2000]{Nishiyama}
Nishiyama M, Hong K, Mikoshiba K, Poo M, Kato K (2000)
Calcium stores regulate the polarity and input specificity of synaptic
modification. {\it Nature} 408:584--588. 

\bibitem[Nowotny et al., 2003]{Nowotny03JNS}
Nowotny T, Zhigulin VP, Selverston AI, Abarbanel HDI,
Rabinovich MI (2003)
Enhancement of synchronization in a hybrid neural circuit
by spike-timing dependent plasticity.
{\it J. Neurosci.} 23(30):9776--9785.

\bibitem[Pikovsky et al., 2001] {Pikovskybook}
Pikovsky A, Rosenblum M, Kurths J (2001)
Synchronization -- A Universal 
Concept in Nonlinear Sciences.
Cambridge University Press, Cambridge, UK.

\bibitem[Plenz and Kitai, 1999]{Plenz99}
Plenz D, Kitai ST (1999)
A basal ganglia pacemaker formed by the subthalamic nucleus and
external globus pallidus. {\it Nature} 400:677--682.  

\bibitem[Ramirez et al., 2004]{Ramirez04}
Ramirez JM, Tryba AK, Pe\~{n}a F (2004)
Pacemaker neurons and neuronal networks: an integrative view.
{\it Curr. Opinion in Neurobiol.} 14:665--674.

\bibitem[Reyes, 2003]{Reyes}
Reyes AD (2003) Synchrony-dependent propagation of firing rate
in iteratively constructed networks {\it in vitro}.  {\it Nature
Neurosci.} 6(6):593--599.

\bibitem[Ritz and Sejnowski, 1997]{Ritz97}
Ritz R, Sejnowski TJ (1997)
Synchronous oscillatory activity in sensory systems:
new vistas on mechanisms.
{\it Curr. Opinion in Neurobiol.} 7:536--546.

\bibitem[Seliger et al., 2002]{Seliger}
Seliger P, Young SC, Tsimring LS (2002)
Plasticity and learning in a network of coupled phase oscillators.
{\it Phys. Rev. E} 65:041906.

\bibitem[Shouval et al., 2002]{Shouval}
Shouval HZ, Bear MF, Cooper LN (2002)
A unified model of NMDA receptor-dependent bidirectional synaptic
plasticity. {\it Proc. Natl. Acad. Sci. USA}, 99(16):10831--10836.

\bibitem[Singer and Gray, 1995]{Singer95}
Singer W, Gray CM (1995)
Visual feature integration and the temporal correlation hypothesis.
{\it Ann. Rev. Neurosci.} 18:555--586.

\bibitem[Song et al., 2000]{Song00}
Song S, Miller KD, Abbott LF (2000) Competitive
Hebbian learning through spike-timing-dependent synaptic
plasticity. {\it Nat. Neurosci.} 3(9):919--926.

\bibitem[Song and Abbott, 2001]{Song01}
Song S, Abbott LF (2001) Cortical development and remapping
through spike timing-dependent plasticity. {\it Neuron}
32:339--350.

\bibitem[Timme et al., 2002]{Timme02}
Timme M, Wolf F, Geisel T (2002)
Prevalence of unstable attractors in networks of pulse-coupled
oscillators. {\it Phys. Rev. Lett.} 89:154105.

\bibitem[van Rossum et al., 2000]{Vanrossum00}
van Rossum MCW, Bi GQ, Turrigiano GG (2000) 
Stable Hebbian learning from spike timing-dependent
plasticity. {\it J. Neurosci.} 20(23):8812--8821.

\bibitem[Vogels and Abbott, 2005]{Vogels}
Vogels TP, Abbott LF (2005)
Signal propagation and 
logic gating in networks of integrate-and-fire neurons.
{\it J. Neurosci.} 25(46):10786--10795.

\bibitem[Winfree, 1980]{Winfree80}
Winfree AT (1980) The Geometry of Biological
Time. Springer-Verlag, New York.

\bibitem[Woodin et al., 2003]{Woodin}
Woodin MA, Ganguly K, Poo M (2003)
Coincident pre- and postsynaptic activity modifies GABAergic
synapses by postsynaptic changes in Cl${}^-$ transporter activity.
{\it Neuron} 39:807--820.

\bibitem[Zhang et al., 1998]{Zhang98}
Zhang LI, Tao HW, Holt CE, Harris WA, Poo M
 (1998) A critical window for cooperation and competition among
developing retinotectal synapses. {\it Nature} 395:37--44.

\bibitem[Zhigulin et al., 2003]{Zhigulin03}
Zhigulin VP, Rabinovich  MI, Huerta R,
Abarbanel HDI (2003)
Robustness and enhancement of neural synchronization by
activity-dependent coupling.
{\it Phys. Rev. E} 67:021901.

\bibitem[Zhigulin and Rabinovich, 2004]{Zhigulin04}
Zhigulin VP, Rabinovich MI (2004)
An important role of spike timing dependent synaptic
plasticity in the formation of synchronized neural ensembles.
{\it Neurocomputing} 58--60:373--378.

\end{thebibliography}
\end{document}